\documentclass[prd,aps,floats,showkeys,nofootinbib,notitlepage]{revtex4-2}
\usepackage{amsmath}
\usepackage{amssymb}
\usepackage{amsbsy}
\usepackage{float}
\restylefloat{table}
\usepackage{graphicx}
\usepackage{epsfig}
\usepackage{epstopdf}
\usepackage{color}
\usepackage[colorlinks=true,linkcolor=red]{hyperref}

\usepackage{listings} 
\usepackage{xcolor} 

\usepackage{tabularx} 
\usepackage{multirow}
\usepackage{enumerate}
\usepackage{relsize}

\allowdisplaybreaks 

\begin{document}

\title{GiNaCDE: the high-performance F-expansion and First
	Integral Methods with C++ library for solving Nonlinear
	Differential Equations\footnote{This article presents a detailed introduction to the NLPDE solver GiNaCDE \cite{ginacde}.}}
\author{Mithun Bairagi}
\email{bairagirasulpur@gmail.com}
\affiliation{Department of Physics, The University of Burdwan, Golapbag 713104, West Bengal, India}

\begin{abstract}
	We present the algorithms for three popular methods: F-expansion, modified F-expansion, and first integral methods to automatically get closed-form traveling-wave solutions of nonlinear partial differential equations (NLPDEs).  We generalize and improve the methods. The proposed algorithms are manageable, straightforward, and powerful tools providing a high-performance evaluation of the exact solutions of nonlinear ordinary differential equations (NLODEs) and NLPDEs. For differential equations with parameters, the new algorithms determine the conditions on the parameters to obtain exact solutions. The algorithms show solutions to a wide variety of NLODEs and NLPDEs, both integrable and non-integrable. It can solve NLODEs and NLPDEs that contain complex functions.
	The algorithms are implemented in a C++ library named GiNaCDE.
	The efficiency and effectiveness of the algorithms are demonstrated by some examples with the help of GiNaCDE. The output results tally with the previously known results, and in some cases, new exact traveling-wave solutions are explicitly obtained. 
	Use of the library, implementation issues, scope, limitations, and future extensions of the software are addressed.

\end{abstract}
\keywords{NLPDEs; F-expansion method; First integral method; Symbolic computations.}
\maketitle

\section{Introduction}
The Nonlinear Ordinary Differential Equations (NLODEs) and Nonlinear Partial Differential Equations (NLPDEs) play an important role in the theoretical sciences to explain many Nonlinear phenomena in various fields of science, such as biology, chemistry, engineer, solid-state physics, plasma physics, optical fibers, and so on. The exact (closed-form) traveling-wave solutions (TWS) of such NLPDEs give much extra information, which helps us to study the result more deeply. The knowledge of closed-form solutions of NLODEs and NLPDEs helps to test the degree of accuracy of numerical solvers and also facilitates stability analysis. In the past few decades, many powerful methods have been presented for seeking exact solutions of NLPDEs. Some of them are tanh-expansion method \cite{tanh,tanh1,tanh2}, Jacobi elliptic function expansion method \cite{jef,jef1}, multiple exp-function method \cite{mulExp}, tanh-sech method \cite{tanhSech}, extended tanh method \cite{extTanh}, hyperbolic function method \cite{hypFunc}, sine–cosine method \cite{sinCos}, inverse scattering method \cite{invScat}, Hirota’s direct method \cite{hirota}, F-expansion method \cite{fexpn}, first integral method \cite{fim0,fim}, etc.
We have observed that most of the methods (such as first seven methods mentioned above) are function-expansion methods where the exact solutions of nonlinear differential equations restrict to polynomial solutions in terms of specific functions. For example the exact solutions are obtained in terms of the tanh function for tanh-expansion method  \cite{tanh,tanh1,tanh2} and Jacobi’s sn (JacobiSN) or cn (JacobiCN) functions for Jacobi-expansion method \cite{jef,jef1}. These methods are useful for finding explicit traveling solitary wave solutions to nonlinear evolution equations. 

On the other hand, F-expansion \cite{fexpn} and first integral methods \cite{fim0,fim} are different kinds of methods, which can overcome the limitations of the function-expansion method. F-expansion method was first introduced by Zhou et al. \cite{fexpn}. Later this method has been further improved \cite{extfexpn1, extfexpn2, extfexpn3,extfexpn4,genfexpn1, genfexpn2,imprfexpn, furimprfexpn,modfexpn}, and successfully applied to many nonlinear equations for finding the exact solutions. In the F-expansion method, solutions of NLPDEs are expressed in finite power series of a new function $F(\xi)$ which depends on traveling-wave coordinate $\xi$. $F(\xi)$ satisfies a first-order NLODE called auxiliary equation (A.E.). By repeated substitutions and using the chain rules arise from A.E., the order of NLPDEs is reduced to first-order NLODEs. As a result, solutions can be easily deduced from known solutions of first-order NLODEs. In the modified F-expansion method \cite{modfexpn} (in short mF-expansion) which is very similar to the F-expansion method, a different kind of A.E. is taken. In our proposed algorithms for F-expansion and modified F-expansion methods, one important advantage is that one can set A.E. in the general forms \eqref{aeF-exp},\eqref{1stnlode2} while in \cite{extfexpn4,modfexpn} the A.E. have been taken in the specific forms \eqref{024}, \eqref{riccati} respectively. Unlike the function-expansion method, a huge variant of solutions can be obtained using different auxiliary equations. Even the solutions obtained from the function-expansion method can also be recovered from the F-expansion method. On the other hand, the first integral method (FIM) is based on the ring theory of commutative algebra. In FIM, a second-order NLPDE is automatically converted to the first-order NLODE when possible, which is known as the first integral of the original second-order NLPDE.  Such automatic conversion is not possible in F-expansion and modified F-expansion methods.  Then the solutions of second-order NLPDE are determined from known solutions of the first-order NLODE. This method has been successfully applied to a number of NLPDEs by researchers \cite{eckhaus,fim1,fim2,fim3,fim4,fim5,fim6,fim7,fim8}. 
Keeping in mind the popularity of the F-expansion, mF-expansion, and first integral methods in the scientific community in finding the exact solutions of NLODEs and NLPDEs, we have been motivated to introduce the algorithms of these methods that can be easily implemented in a computer package for automatic derivations of closed-form traveling-wave solutions by employing these methods. In our proposed algorithms, we have generalized these methods and introduced some new features in the methods so that one can apply the new algorithms to different kinds of NLODEs and NLPDEs arising in various fields of science.    

Searching for traveling-wave solutions of NLPDEs by hand is very tedious, cumbersome, and sometimes takes a long time. However, many modern Computer Algebra Systems (CAS), such as Maple, Mathematica, Reduce, Maxima, etc., help us to perform complicated and tedious algebraic calculations on a computer to find exact solutions to such problems accurately in a short time.
In order to solve the NLPDEs, many computer packages are available.
In 1996, Parkes and Duffy \cite{atfm} have implemented tanh-expansion in their Mathematica package ATFM. Later complete implementation of tanh-expansion has been done by Li and Liu (2002) \cite{rath} designing the Maple package RATH. Baldwin et al. \cite{pdespclpkg, pdespclpkg1} have developed Mathematica package { \em PDESpecialSolutions.m} which admits polynomial solutions in tanh, sech, combinations thereof, JacobiSN, JacobiCN. RAEEM \cite{raeem} is one of the most popular packages written in the Maple programming language, which is a comprehensive and complete implementation of some powerful methods such as the tanh-method, the extended tanh-method, the Jacobi elliptic function method, and the elliptic equation method. All these computer packages have implemented the function-expansion methods. One serious drawback of the function-expansion method is that the solutions which contain functions other than some specific type of functions, such as tanh, sech, JacobiSN, JacobiCN, etc. are not obtained. Additionally, the non-polynomial forms of these specific functions are not obtained also.
We note that most packages have been developed within commercially available software frameworks, such as Maple and Mathematica.  
Besides this, to the best of our knowledge, the computer packages with the implementation of F-expansion and first integral methods are not available so far.

Keeping in mind all the above points of view, we have been motivated to develop a computer package or C++ library called GiNaCDE \footnote{\url{https://github.com/mithun218/GiNaCDE.git}} \cite{ginacde}, which implements our proposed algorithms of the high-performance automated F-expansion, modified F-expansion, and first integral methods for the first time. The symbolic manipulations of GiNaCDE depend only on GiNaC \cite{ginac}. GiNaCDE extends the capability of the GiNaC library towards the solutions of differential equations. There are several advantages to use GiNaC over other CAS. GiNaC is a pure C++ library. It can accept C++ programming language, a general-purpose object-oriented programming (OOP) language, and it is fast like commercially available computer algebra systems. The source code is freely available as this library is written under the terms and conditions of the GNU general public license (GPL). One important property of GiNaC that differentiates it from other computer algebra programs we may have used: GiNaC assigns a unique (hidden) serial number for each newly created symbol object, and GiNaC uses this unique serial number instead of its name for algebraic manipulations. The serial number for the same name of a symbol may be changed in each running session of the GiNaC program. As a result, the symbols in the same algebraic expressions may be ordered differently during each running session of the GiNaC program. This happens because to order the symbols of an algebraic expression GiNaC internally uses a comparison predicate, called {\em ex\_is\_less} which uses an internal symbol id counter. There is also have another problem with using GiNaC as CAS.  One should write a C++ program every time to solve each NLPDE, and compilation is also required for each one. However, many GUI tools like GTK, FLTK, Qt are available to make GUI applications with GiNaCDE, which facilitates users to solve NLPDEs automatically without writing programming and compilation each time. In this context, we have developed a GUI application called GiNaCDE GUI within the GTK framework. 

We have noted that in the Maple package RAEEM \cite{raeem}, similar to the F-expansion method, an A.E. has also been considered, whose algebraic form is very similar to the A.E. as taken in the F-expansion method. However, in RAEEM, A.E. is handled with a maximum of four parameters, but in the case of GiNaCDE, one can choose the number of parameters higher than four that can be controlled from the user end. In order to get the exact traveling wave solutions including polynomial, exponential, triangular, hyperbolic, rational, Jacobi elliptic, Weierstrass elliptic type, RAEEM automatically chooses some predefined constant values of the four parameters of A.E., and these values cannot be controlled from the user end. On the other hand, in GiNaCDE software, one can assign any numerical or algebraic values to the parameters of A.E. and choose the upper limit on the number of parameters in A.E. in own choices. All this flexibility in handling the parameters of A.E. from the user end makes the GiNaCDE software more powerful and efficient to investigate more new types of closed-form solutions of NLPDEs.

The paper is organized as follows: In Secs. \ref{sec:Fexpn}, \ref{sec:mF}, \ref{sec:fim}, we have simply described the algorithms for automated F-expansion, modified F-expansion, and first integral methods respectively and described each step of the algorithms with a working example of NLPDE. Key features of the implementation of the proposed algorithms in GiNaCDE software are discussed in Sec. \ref{sec:imple}.
In Sec. \ref{sec:examp}, to illustrate the efficiency and effectiveness of the GiNaCDE library, we have applied the library to a wider class of NLPDEs and found the exact solutions. Our results are also compared with an existing Maple package RAEEM.  Finally, in Sec. \ref{sec:conc}, we discuss our results, and some conclusions have been made on our works. The exact analytical solutions of some auxiliary equations are given in the Appendix: \ref{sec:appenA}, \ref{sec:appenB}, \ref{sec:appenC}, \ref{sec:appenD}. 

\section{Algorithm of automated F-expansion method}\label{sec:Fexpn}
This section explains the algorithms for the F-expansion method \cite{extfexpn4} to obtain closed-form traveling-wave solutions of NLPDEs automatically. However, in the proposed algorithms, we have to guide the solution process initially by providing some initial data. In that sense, the algorithms are not fully automated, but this makes the algorithms more powerful, and we can apply these algorithms to a huge variant of NLPDEs with different types of initial data.
We divide the algorithm into five main steps (labeled F1-F5).  
\par Let us consider an NLPDE with independent variables $t,x_1,x_2, \ldots x_m=\mathbf{X}$ and dependent variable $u$ in the following general form
\begin{equation}\label{geneq}
	F\left( {\alpha_i,u,u_t,u_{x_1},u_{x_1} \ldots ,u_{x_m},u_{tt},u_{t{x_1}},u_{t{x_2}}, \ldots ,u_{t{x_m}},u_{{x_1}{x_1}},u_{{x_1}{x_2}} \ldots ,u_{{x_1}{x_m}} \ldots } \right) = 0,
\end{equation}
where $\alpha_i(i=1,\ldots,l)$ are the parameters, $u=u(t,x_1,x_2, \ldots x_m)$ and $F$ is a polynomial about $u$ and its derivatives. Equation \eqref{geneq} does not explicitly depend on the independent variables $t,x_1,x_2, \ldots x_m$. The current package GiNaCDE with the all three algorithms (F-expansion, mF-expansion, FIM) can solve the NLPDEs of the form \eqref{geneq}. However, there is no guarantee that the code always give the complete solutions of all NLPDEs of the form \eqref{geneq}, and sometimes the code may failure to give solutions due to the complexity in the problems. Here, it should be noted that our algorithms are also applicable to Eq. \eqref{geneq} without the parameters $\alpha_i$. Equation \eqref{geneq} is called input-NLPDE throughout this paper.
\\
\textbf{Step F1}(Transform the NLPDE into a NLODE): At first we take a transformation in Eq. \eqref{geneq}
\begin{equation}\label{twtrans}
	u = U(\xi )e^{I\theta},
\end{equation}
where
\begin{align}
	\label{twco}& \xi = k_0t + k_1{x_1} + {k_2}{x_2} +  \ldots  + {k_m}{x_m}=\boldsymbol{K}\cdot\mathbf{X},\nonumber\\
	\intertext{and}& \theta = {p_0}t + {p_1}{x_1} + {p_2}{x_2} +  \ldots  + {p_m}{x_m}=\boldsymbol{P}\cdot\mathbf{X}.
\end{align}
Here $\xi$ is called traveling-wave coordinate and $U(\xi )$ is the traveling-wave part  of the solutions. The second part of Eq. \eqref{twtrans} is $e^{I\theta}$ which is called the phase part of the solutions and $\theta$ is called phase coordinate. Usually, this part is present when differential equation \eqref{geneq} has an imaginary part. In our program library we must retain this phase part when Eq. \eqref{geneq} is complex otherwise we only retain traveling-wave part. Taking proper forms or values of the constant coefficients ($k_i\;(i=0\ldots m)$) of traveling-wave coordinate $\xi$ and phase angle constants $p_i\;(i=0\ldots m)$, we can transform the NLPDE \eqref{geneq}  into the nonlinear ordinary differential equation (NLODE) with single independent variable $\xi$ and dependent variable $U(\xi)$. Repeatedly applying the chain rule 
\begin{equation}\label{twtrans2}
	\frac{\mathrm{\partial} \bullet }{\mathrm{\partial} \mathbf{X}}=\left ( \boldsymbol{K}\frac{\mathrm{d} }{\mathrm{d} \xi}+I\boldsymbol{P} \right )\bullet
\end{equation}
on Eq. \eqref{twtrans}, Eq. \eqref{geneq} is transformed into NLODE. The form of NLODE in general form is given by
\begin{equation}\label{tweq}
	G\left( {\alpha_i,k_i,p_i,U,U^{(1)},U^{(2)}} \ldots \right) = 0,
\end{equation}
where $U^{(n)} (n=1,2\ldots)$ indicates $n$ times differentiation with respect to $\xi$. All the three algorithms (F-expansion, mF-expansion, FIM) work only when Eq. \eqref{tweq} is a polynomial in variable $U(\xi)$ and its derivatives, and does not explicitly depend on the independent variables $\xi$. Then our algorithm check the integrability of Eq. \eqref{tweq}. If Eq. \eqref{tweq} is integrable, our method try to integrate Eq. \eqref{tweq} and if integration is successful then one can assign a value to each integrating constant ($ic_i,\;i=1,2,\ldots,\eta$, $\eta$ is the number of integration) in own choices. In the case of complex NLPDE, if real and imaginary both parts are present in the transformed NLPDE, one part is taken to solve. To select one part from complex NLODE, following strategies are followed
\begin{itemize}
	\item[i.] Assume all the parameters $(\alpha_i,k_i,p_i)$ are real.
	\item[ii.] Express the NLODE \eqref{tweq} in the form 
	\begin{equation}\label{tweq3}
		G\left( {\alpha_i,k_i,p_i,U,U^{(1)},U^{(2)}} \ldots \right) = Re\left( {\alpha_i,k_i,p_i,U,U^{(1)},U^{(2)}} \ldots \right)+Im\left( {\alpha_i,k_i,p_i,U,U^{(1)},U^{(2)}} \ldots \right)*I=0,
	\end{equation}
	where $Re\left( {\alpha_i,k_i,p_i,U,U^{(1)},U^{(2)}} \ldots \right)$ is real part and $Im\left( {\alpha_i,k_i,p_i,U,U^{(1)},U^{(2)}} \ldots \right)$ is imaginary part of NLODE \eqref{tweq3}, and $I=\sqrt{-1}$.
	\item[iii.] Check in which part ($Re$ or $Im$) dependent variable $U(\xi)$ is not present, but minimum one number of parameter is present there. Get the constraint on that parameters contained in this part. Take other part (may be $Re$ or $Im$) as NLODE whose solutions are to be solved. 
	\par Suppose in the real part $Re$, the dependent variable $U(\xi)$ is not present and there is only present the parameters $\alpha_i,k_i,p_i$. Then Eq. \eqref{tweq3} can be expressed by 
	\begin{equation}
		Re\left( {\alpha_i,k_i,p_i} \right)+Im\left( {\alpha_i,k_i,p_i,U,U^{(1)},U^{(2)}} \ldots \right)*I=0,
	\end{equation}
	and we have to solve the NLODE $Im\left( {\alpha_i,k_i,p_i,U,U^{(1)},U^{(2)}} \ldots \right)=0$ subject to the constraint $Re\left( {\alpha_i,k_i,p_i}\right)=0$.
	\item[iv.] If the above step is failure, compare the expressions of $Re$ and $Im$ to check whether they are the same NLODE for some constraint on the parameters of $Re$ or $Im$. If they are the same NLODE, take anyone part as NLODE whose solutions are to be solved.
	\par Suppose for the constraint 
	\begin{equation}\label{consf}
		f(\alpha_i,k_i,p_i)=0,
	\end{equation}
	we get
	\begin{equation}\label{reEqIm}
		Re\left( {\alpha_i,k_i,p_i,U,U^{(1)},U^{(2)}} \ldots \right)=Im\left( {\alpha_i,k_i,p_i,U,U^{(1)},U^{(2)}} \ldots \right)
	\end{equation}
	from Eq. \eqref{tweq3}, then the solutions can be determined either from the NLODE $Re$ or $Im$ subject to the constraint relation \eqref{consf}. 
\end{itemize}
If any above criteria are not satisfied, the complex input-NLPDE equation cannot be transformed into single NLODE in our algorithm and the method does not work
successfully.\\

\textbf{Step F2}(Determine the highest power $N$ of finite power series): 
Now according to the F-expansion method, the solution of the NLODE \eqref{tweq} can be expressed as the finite power series, which is
\begin{equation}\label{soluseries}
	U =  \sum\limits_{i = 0}^N {a_i{F^i}(\xi )}+\sum\limits_{i = 1}^N {\frac{{{b_i}}}{{F^i}(\xi )}},
\end{equation}
where $a_i(i=0\ldots N)$ and $b_i(i=1\ldots N)$ are constants to be determined later.
In Eq. \eqref{soluseries} first term is positive part and second term is negative part in the solution. The value of $N$ (positive integer or positive noninteger) in Eq. \eqref{soluseries} can be determined by considering a homogeneous balance between the highest order nonlinear term with the highest order derivative of $U(\xi)$ in Eq. \eqref{tweq}.
However, to automate this process, we have employed a method described in \cite{rath}. At first, Eq. \eqref{tweq} is expanded in the sum of product (SOP) form. Our aim is to determine the highest possible value of $N$, and so it is sufficient to replace $U$ with $U^N$. Assuming degree of $U(\xi)$ is $D[U(\xi)]=N$, we replace $U$ by $U^N$ and collect the degrees of each term appearing in Eq. \eqref{tweq} by a variable, say {\em E}. To determine the degree of an expression, we use the relations
\begin{equation}\label{degRltn}
	D\left[ {\frac{{{d^n}U(\xi )}}{{d{\xi ^n}}}} \right] = N + n,\,\,D\left[ {{U^n}{{\left( {\frac{{{d^n}U(\xi )}}{{d{\xi ^n}}}} \right)}^m}} \right] = Nn + m(N + n).
\end{equation}
Generally in ({\em max( E)},$N$) plane we get a turning point and the value of $N$ is taken at this point. Whole procedure is automated in the following order:
\begin{itemize}
	\item[i.] Substitute $U$ by $U^N$ and simplify.
	\item[ii.] Expand and express in SOP form.
	\item[iii.] Collect the degree of each term appearing in SOP form and store them in a list $ E$.
	\item[iv.] Replace $N$ in $E$ by a sequence of numbers whose first number is 0. In GiNaCDE, we take three  number sequence with common differences- $1/2,1/3,1/4$ up to last number 11.
	\item[v.] For each number in sequence, calculate $ max( E)$.
	\item[vi.] Calculate differences between the value of  $ max( E)$ for successive numbers in the sequence.
	\item[vii.] To get highest power $N$, take the number in sequence for which differences are not same with previous one. More clearly if $i$ is the current number in the the sequence and $ ({max(E_i)} - {max(E_{i - 1})}) \ne ({max(E_{i - 1})} - {max(E_{i - 2})})$, then the highest power $N=i-1$.   
\end{itemize}
It is clear that for a larger value of $N$ the complexity in mathematical operations is increased. To avoid such complexity in derivations, one should set the maximum allowed value of $N$. However, sometimes in some cases, the auto-evaluation of $N$ may fail. \\

\textbf{Step F3}(Derive the system of Nonlinear Algebraic Equations for the coefficients of $F(\xi)$):
$F(\xi)$ satisfy the first-order nonlinear ODE (also called auxiliary equation (A.E.))
\begin{equation}\label{1stnlode}
	{F'\left( \xi  \right)} = \mathcal{F}(F(\xi)),
\end{equation}
where $\mathcal{F}(F(\xi))$ is some known functions of $F(\xi)$. The prime over $F(\xi)$ represents differentiation with respect to $\xi$. In case of F-expansion method
\begin{equation} \label{aeF-exp}
	\mathcal{F}(F(\xi)) = \sqrt{{A_0} + {A_1}F + {A_2}{F^2} +  \ldots + {A_\delta }{F^\delta }},
\end{equation}
where $\delta$ is a positive integer and $A_i(i=0,1,\ldots \delta)$ are coefficients of A.E. Here one can use any functional form of Eq. \eqref{aeF-exp} by choosing any positive integer value of $\delta$ and real values of $A_i$. As a result the solutions of first-order NLODE \eqref{aeF-exp} can be expressed in terms of a large variety of functions such as polynomial, exponential, trigonometric, hyperbolic, rational, Jacobi elliptic etc.. 
However, taking some well-known functional forms of $\mathcal{F}$ in Eq. \eqref{aeF-exp}, we have shown some solutions of $F(\xi)$ in the Appendix: \ref{sec:appenC}, \ref{sec:appenD}. The higher derivatives of $F(\xi)$ using Eq. \eqref{1stnlode} can be expressed by
\begin{equation}\label{1stnlodeD}
	F^{\prime\prime}= \frac{\mathrm{d} \mathcal{F} }{\mathrm{d}F}\mathcal{F},\;\;F^{\prime\prime\prime}=\left(\frac{\mathrm{d} \mathcal{F} }{\mathrm{d}F}\mathcal{F}\right)^2+\mathcal{F}^2\frac{\mathrm{d^2} \mathcal{F} }{\mathrm{d}F^2}\;\; \text{and so on.}
\end{equation}
Now substituting Eq. \eqref{soluseries} into Eq. \eqref{tweq} and using Eq. \eqref{1stnlodeD} with \eqref{aeF-exp} we get an expression, and the numerator of resulted expression contains $F(\xi)^j{\mathcal{F}}^k\;(j=0,1,2,\ldots;k=0,1)$ terms. Setting each coefficient of $F(\xi)^j{\mathcal{F}}^k$ to zero an overdetermined system of nonlinear algebraic equations are obtained where the constant parameters $a_i(i=0\ldots N)$, $b_i(i=1\ldots N)$, $k_i(i=0\ldots m)$, $p_i(i=0\ldots m)$, $A_i(i=0,1,\ldots \delta)$, parameters $\alpha_i(i=1\ldots l)$ appearing in input-NLPDE and integrating constants $ic_i(i=1\ldots \eta)$ if \eqref{tweq} is integrable, are present. 

The nonlinear system of equations is a set of simultaneous equations in which the unknowns (the constant parameters) appear as variables of a polynomial of degree one or higher than one.
Suppose the system of algebraic equations is solved for all parameters that are present in the system of algebraic equations. In that case, it takes a larger time to get solutions for the system of equations, even sometimes solutions are not obtained for a complicated system. To reduce the calculating time and the complexity in derivations, we categorize all the parameters into two different types. They are external parameters and internal parameters. External parameters are $\alpha_i(i=1\ldots l)$, $A_i(i=0,1,\ldots \delta)$ which are present in input-NLPDE equation and auxiliary equation respectively. When input-NLPDE equations are integrable, the generated integrating constant(s) $ic_i(i=1\ldots \eta)$ is also external parameter. On the other hand all the remaining parameters, such as $a_i(i=0\ldots N)$, $b_i(i=1\ldots N)$, $k_i(i=0\ldots m)$, $p_i(i=0\ldots m)$ are called internal parameters as they generate internally. Nonlinear algebraic systems are always solved for internal parameters. But one can have control over external parameters to choose the parameters for which nonlinear algebraic system is to be solved. For this purposes, the programming variables {\em ASolve} and {\em paraInDiffSolve} (detailed descriptions are given in the software user manual) are used to choose the unknowns from the external parameters in our choice, and it will reduce the calculating time and can handle more complicated algebraic expressions. At the same time, the exact solutions are determined subject to the conditions on chosen external parameters.    \\

\textbf{Step F4}(Solve the system of Nonlinear Algebraic Equations): Analysing and solving the nonlinear algebraic system is a vital and challenging step among all steps of the method. In fact, the number of exact solutions of NLPDE derived by F-expansion, modified F-expansion, or first integral methods is entirely depending on how many solutions are obtained from a nonlinear algebraic system. The executing time of the software mainly depends on this step. Many methods are available to solve nonlinear algebraic systems such that Gröbner basis methods \cite{grobner}, the Ritt-Wu characteristic sets method implemented by Wang \cite{RittWu,RittWu1}, and the Reduced Involutive Form (Rif) code by Wittkopf and Reid \cite{rif}. D. Baldwin et al. \cite{baldwin} have employed a simple algorithm to design a powerful nonlinear solver in Mathematica. We have followed their algorithm to create a nonlinear solver in GiNaC symbolic system. The nonlinear solver implemented in \cite{baldwin} solves the entire system in an automated way using the built-in Mathematica function \textit{Reduce}. Their solver can solve polynomial and non-polynomial systems both. The nonlinear solver implemented by us can solve only the polynomial system required in this application, and its own C++ function solves the polynomial equations.  

The steps used in this algorithm are very much like the steps used to solve a nonlinear algebraic system by hand. In this method, the simplest equation is solved for sorted unknown parameters. Then the solutions are substituted in the remaining equations. Such solving and substitution procedures are repeated until the system is completely solved. We operate the whole procedure in the following order:
\begin{itemize}
	\item[i.] Check whether each equation is polynomial in unknowns.
	\item[ii.] Factor and simplify each equation.
	\item[iii.] Measure complexity of each equation by the number of add containers, unknown parameters, and the degree of unknowns. Then, sort the system based on their complexity. If more than one equations have the same complexity, in GiNaCDE they are sorted according to the GiNaC in-built comparison predicate {\em ex\_is\_less}.
	\item[iv.] Sort the unknown parameters contained in the simplest equation by their degree.
	\item[v.] Solve the simplest equation for the lowest degree unknown. If the number of unknown for the lowest degree is greater than one, then GiNaCDE uses the comparison predicate {\em ex\_is\_less} to choose the unknown. If solutions are absent, solve the unknown for the next higher degree. 
	\item[vi.] Substitute the solutions into the remaining equations and simplify.
	\item[vii.] Repeat the steps i-vi until all the equations are reduced to zero.
	\item[viii.] Substitute all the unknowns which are present in the computed solutions with the help of other solutions.
	\item[ix.] Test the solutions by substituting them into each equation.
	\item[x.] Finally, collect all solutions branches.
\end{itemize}
Our solver is powerful and can easily handle nonlinear equations (of course, polynomials in unknowns) with multi-parameters. Sometimes, there are risks of missing some solutions due to numerous parameters in the system or if the system is high degree. In this solver, the unknowns from all parameters appearing in the system are chosen in order of complexity. Then the solutions for these unknowns are expressed in terms of other parameters that are to be regarded as arbitrary parameters. Sometimes it is observed that the solutions become simpler where these arbitrary parameters are taken as unknowns.\\

\textbf{Step F5}(Build solutions with calculating steps): Substitute the solutions obtained in step F4 into Eq. \eqref{aeF-exp}  and obtain the solutions of $F$ using the Appendix: \ref{sec:appenC}, \ref{sec:appenD}. Then, to obtain traveling-wave solutions of Eq. \eqref{tweq}, substitute $F$ and the solutions obtained in step F4 into Eq. \eqref{soluseries}. Finally the explicit solutions in original variables are obtained using Eqs. \eqref{twtrans}, \eqref{twco}.\\

\subsection{Explanation of each step with a working example}
To illustrate each algorithm steps for F-expansion method described above, we take the one dimensional cubic nonlinear Schr\"odinger (NLS) equation \cite{nlse}
\begin{equation}\label{nls}
	Iu_t-pu_{xx}+q{|u|}^2u=0,
\end{equation}
as an example. In Eq. \eqref{nls} $p,q$ are non-zero real constants and $u(x,t)$ is a complex-valued function depends on the variables $x,t$.

\textbf{Step F1:} Equation \eqref{nls} has an imaginary and a real part. So retaining the phase part in Eq. \eqref{twtrans} and making traveling-wave transformation
\begin{equation}\label{nls_twtrans}
	u(t,x)=U(\xi)e^{Ip_0t+Ip_1x},\;\;\;\text{where}\;\;\xi=k_0t+k_1x,
\end{equation}
Eq. \eqref{twtrans2} yields 
\begin{align}\label{nlstwtrans}
	&u_t=k_0 U_\xi+Ip_0,\nonumber\\
	&u_{xx}=k_1^2U_{\xi\xi}+2Ik_1p_1U_\xi-p_1^2.
\end{align}
Substituting Eqs. \eqref{nlstwtrans} into Eq. \eqref{nls} we get
\begin{equation}\label{nlsbothpart}
	- {p_0}U - pk_1^2{U_{\xi \xi }} + q{U^3} + p_1^2pU+IU_\chi\left({k_0} - 2{p_1}p{k_1}\right)=0.
\end{equation}  
Note that, Eq. \eqref{nlsbothpart} has real and imaginary both parts. From imaginary part we get the condition
\begin{equation}\label{nlscond}
	{p_1} = \frac{{{k_0}}}{{2p{k_1}}},
\end{equation}
and from real part we obtain the NLODE 
\begin{equation}\label{nlsNLODE}
	- pk_1^2{U_{\xi \xi }} + q{U^3}- {p_0}U=0.
\end{equation}

\textbf{Step F2:} Substituting $U\rightarrow U^N$, degree of each term in Eq. \eqref{nlsNLODE} is collected by a list variable {\em E}:
\begin{equation}
	E=\left(D[U_{\xi \xi }],D[U^3],D[U]\right)=(N+2,3N,N).
\end{equation}
\begin{table}[H]
	\centering
	\caption{Here a number sequence with common difference $1/2$ has been used to determine the highest power $N$. Bold number denotes the turning point and the corresponding value $N=1$ is balancing highest power. }
	\begin{tabular}{c@{\hskip .0in}c@{\hskip .1in}c@{\hskip .1in}c@{\hskip .1in}c}
		& $N$           & $max(E_i)$ & $max(E_i)-max(E_{i-1})$\\ [0.5ex]
		\hline \hline\\
		& $0$ & $2$      & $ $                    \\[0.5ex]
		& $\frac{1}{2}$ & $2.5$      & $0.5$                    \\[0.5ex]
		&  $1$          & $3$        & $\mathbf{0.5}$                            \\[0.5ex]
		&  $\frac{3}{2}$& $4.5$      & $1.5$                  \\[0.5ex]
		&  $2$          & $6$        & $1.5$ \\
		[1ex] 
		\hline 
	\end{tabular}
	\label{tab:phicompare}
\end{table}

Then using balancing highest power $N=1$, finite power series is expressed by
\begin{equation}\label{nlspowseries}
	U= a_0+a_1F+{b_1}{F^{-1}}.
\end{equation}

\textbf{Step F3:} We choose A.E.
\begin{equation}\label{nlsAE}
	F^{\prime}=\sqrt{A_0+A_2F^2}.
\end{equation}
Using Eqs. \eqref{1stnlodeD} for Eq. \eqref{nlsAE}, we get 
\begin{equation}\label{1stnlodeDnls}
	F^{\prime\prime}=A_2F,\;F^{\prime\prime\prime}=A_2F^{\prime}.
\end{equation}
Inserting Eqs. \eqref{nlspowseries}, \eqref{nlsAE}, \eqref{1stnlodeDnls} into Eq. \eqref{nlsNLODE} and collecting the coefficients of $F^iF^{\prime^j}$ $(i=0,1,\ldots 6,j=0)$ we get a set of nonlinear algebraic equations
\begin{align}
	&F^0:\;\;-8A_0p^2k_1^4b_1+4qpk_1^2b_1^3=0,\nonumber\\
	&F^1:\;\;12qpa_0k_1^2b_1^2 = 0,\nonumber\\
	&F^2:\;\; b_1k_0^2-4p^2k_1^4A_2b_1+12qa_1pk_1^2b_1^2-4pp_0k_1^2b_1+12qpa_0^2k_1^2b_1 = 0,\nonumber\\
	&F^3:\;\; 4qpa_0^3k_1^2+24qa_1pa_0k_1^2b_1+a_0k_0^2-4pa_0p_0k_1^2 = 0,\nonumber\\
	&F^4:\;\; -4a_1p^2k_1^4A_2+12qa_1pa_0^2k_1^2+12qa_1^2pk_1^2b_1+a_1k_0^2-4a_1pp_0k_1^2 = 0,\nonumber\\
	&F^5:\;\; 12qa_1^2pa_0k_1^2 = 0,\nonumber\\
	&F^6:\;\; 4qa_1^3pk_1^2 = 0.
\end{align}
Here external parameters are $A_0,A_2,p,q$ and internal parameters are $a_0,a_1,b_1,k_0,k_1,p_0,p_1$.\\

\textbf{Step F4:} Resulting nonlinear algebraic equations are solved for internal parameters only and the solutions are
\begin{subequations}\label{nlssolu}
	\begin{align}
		\label{nlssolu1}&{a_1} = 0,{k_1} = \pm\sqrt {\frac{q}{{2p{A_0}}}} {b_1},{p_0} = \frac{{A_0^2k_0^2 - {q^2}{A_2}b_1^4}}{{2{A_0}qb_1^2}},{a_0} = 0;\\
		\label{nlssolu2}& { a_1}=0,{ b_1}=0,{ p_0}={\frac {4\,qp{{a_0}}^{2}{{ k_1}}^{2}+{{ k_0}}^{2}}{4p{{ k_1}}^{2}}}.
	\end{align}
\end{subequations}
In the above solutions the parameters $p,q,k_0,p_0,A_0,A_2$ are arbitrary.

\textbf{Step F5:} The solutions for $F$ of Eq. \eqref{nlsAE} are derived with the help of Eq. \eqref{02_solu} in Appendix: \ref{sec:appenD}. Then
combining \eqref{nlspowseries} with \eqref{nlscond} and substituting \eqref{nlssolu1}, we obtain the closed-form solutions to \eqref{nls}. For brevity the solutions are not shown here. Here we present only one solution which illustrates our algorithm:
\begin{align}
	&u(t,x) =  \frac{{\left( { - {A_0} {{\text{e}}^{2 \xi \sqrt {{A_2}} }} + {{\text{e}}^{2 C\sqrt {{A_2}} }}} \right){{\text{e}}^{ - \left( {C + \xi } \right)\sqrt {{A_2}} }}}}{{2\sqrt {{A_2}} }}{{ e}^{I\theta}}, 
\end{align}
where $\theta=\left( {\frac{{A_0^2k_0^2 - {q^2}{A_2}b_1^4}}{{2{A_0}qb_1^2}}t \pm \frac{{\sqrt {{A_0}} {k_0}}}{{\sqrt {2pq} {b_1}}}x} \right), \xi=k_ot\pm \sqrt {\frac{q}{{2p{A_0}}}} {b_1} x$ and $C$ is an arbitrary constant. There are two sign combinations in each equation and total 12 solutions are obtained.  Similarly, using \eqref{nlssolu2}, we can derive an another exact solution
\begin{equation}
	u(t,x)={ a_0}{{ e}^{I\theta}},\;\;\text{where}\; \theta=\frac{{\left( {4 qp{a_0}^2{k_1}^2 + {k_0}^2} \right)t}}{{4p{k_1}^2}} + \frac{{{k_0}x}}{{2p{k_1}}}.
\end{equation}

\section{Algorithm of automated modified F-expansion method}\label{sec:mF}
The algorithm for the automated modified F-expansion method \cite{modfexpn} is very similar to the algorithm of the automated F-expansion method. This algorithm also has five main steps (labeled MF1-MF5), and it has only one difference to the F-expansion method. The difference is that a different form of A.E. is taken in step MF3 in comparison to step F3. Therefore, one can check new exact solutions of the NLPDE applying both methods (F-expansion method and modified F-expansion method) to the same NLPDE with different forms of A.E. Details of all steps are described below:\\ 

\textbf{Step MF1}(Transform the NLPDE into an NLODE): Same as step F1.

\textbf{Step MF2}(Determine the highest power $N$ of finite power series):  Same as step F2.

\textbf{Step MF3}(Derive the system of Nonlinear Algebraic Equations for the coefficients of $(F(\xi)$):
In the modified F-expansion method, the solution of the NLODE \eqref{tweq} is also expressed by a finite series like Eq. \eqref{soluseries}. In this method, we have generalized the modified F-expansion method \cite{modfexpn} taking the A.E. in more general form
\begin{equation}\label{1stnlode2}
	F'\left( \xi  \right) = {A_0} + {A_1}F + {A_2}{F^2} +  \ldots + {A_\delta }{F^\delta },
\end{equation}
where $\delta$ is a positive integer and $A_i(i=0,1,\ldots \delta)$ are coefficients of A.E. One can choose any functional form of Eq. \eqref{1stnlode2} using any positive integer value of $\delta$ and any real value of $A_i$. Here interestingly, we note that by choosing various functional forms of Eq. \eqref{1stnlode2} in our choices, one can get the final solutions of input-NLPDE in terms of a large variety of functions. For example, some well-known equations can be obtained from Eq. \eqref{1stnlode2}, such as Riccati equation with $\delta=2$ and Bernouli equation with $A_i=0,\;(i\ne 1\; \text{and}\; i \ne \delta)$. The exact solutions of Riccati and Bernouli equations are known that are given in Appendix: \ref{sec:appenA} and Appendix: \ref{sec:appenB} respectively.
Now substituting Eq. \eqref{soluseries} into Eq. \eqref{tweq} and using Eq. \eqref{1stnlode2} we get an expression appearing the terms $F(\xi)^j\;(j=0,1,2,\ldots)$ in the numerator. The equations must vanish identically. Hence, to generate a nonlinear algebraic system, equate to zero the coefficients of the power terms in $F$.

\textbf{Step MF4}(Solve the system of Nonlinear Algebraic Equation): Similar strategy as in step F4.

\textbf{Step MF5}(Build solutions with calculating steps): Substitute the solutions of step MF4 into Eq. \eqref{1stnlode2}. Obtain the solutions of $F$ using Appendix: \ref{sec:appenA}, \ref{sec:appenB}. Then, substitute $F$ along with the solutions of step MF4 into Eq. \eqref{soluseries}. To get the explicit solutions in original variables, Eqs. \eqref{twtrans}, \eqref{twco} are used.

As the modified expansion method is very much similar to the F-expansion method (the difference with the F-expansion method is that a different form of A.E. is taken here), we do not explain each step of this method with a working example.  

\section{Algorithm of automated first integral method}\label{sec:fim}
In first integral method \cite{fim0}, one important advantage over F-expansion and modified F-expansion methods is that one does not have to choose A.E. to solve NLPDEs; instead, the input-NLPDE is automatically reduced to a suitable first-order NLODE whose solutions have to be calculated.  

The algorithm for the automated first integral method has eight main steps (labeled FIM1-FIM8). Now we give an outline of every step as follows:\\

\textbf{Step FIM1}(Transform the NLPDEs into NLODEs): Same as step F1. The condition for applying first integral method to the Eq. \eqref{tweq} is that Eq. \eqref{tweq} must be a second-order NLODE. Therefore Eq. \eqref{tweq} is expressed in the form
\begin{equation}\label{tweq2}
	\mathbf{G} \left( {\alpha_i,k_i,p_i,U,U^{(1)},U^{(2)}} \right) = 0.
\end{equation}

\textbf{Step FIM2}(Convert into a system of NLODEs): We assume that $U(\xi)=X(\xi)$ and introducing a new independent variable $Y(\xi) = X_{\xi}(\xi)$, Eq. \eqref{tweq2} can be rewritten as a system of NLODEs \cite{fim0}
\begin{subequations}\label{sysode}
	\begin{align}
		&X_{\xi}(\xi)= Y(\xi),\label{sysode1}\\
		&Y_{\xi}(\xi)= \frac{P(X(\xi),Y(\xi))}{H(X)} = \frac{1}{H(X)}\left(K_0(X)+K_1(X)Y+\ldots+K_d(X)Y^d\right)\label{sysode2}.
	\end{align}
\end{subequations}
We have expressed $P(X(\xi),Y(\xi))$ as a polynomial in variable $Y(\xi)$ with degree $d$ and $H(X),K_i(X)(i=0,1\ldots d)$ are polynomials in variable $X$.
$H(X)$ is the coefficient of the highest derivative term in Eq. \eqref{tweq2}.\\

\textbf{Step FIM3}(Apply Division Theorem): If $X(\xi),Y(\xi)$ are nontrivial solutions of Eq. \eqref{sysode}, then applying the Division Theorem \cite{fim0} there exist an irreducible polynomial in the complex domain $C[X ,Y ]$ such that
\begin{equation}\label{irred}
	q(X(\xi),Y(\xi))=\sum\limits_{i = 0}^N {{a_i}(X){Y^i}}=0,
\end{equation}
where $a_i(i=0\ldots N)$ are polynomials of $X$ and $a_N\neq 0$. Equation \eqref{irred} is called the first integral to Eqs. \eqref{sysode1} and \eqref{sysode2}. Using Division Theorem there exists a polynomial $(g(X)+h(X)Y)$ such that
\begin{equation}\label{qform}
	\frac{{dq}}{{d\xi }} = \frac{{\partial q}}{{\partial X}}\frac{{dX}}{{d\xi }} + \frac{{\partial q}}{{\partial Y}}\frac{{dY}}{{d\xi }} = \left( {g(X) + h(X)Y} \right)\sum\limits_{i = 0}^N {{a_i}(X){Y^i}}.
\end{equation}
Using Eqs. \eqref{sysode}, \eqref{irred} in the Eq. \eqref{qform}, we get
\begin{equation}\label{qred}
	\sum\limits_{i = 0}^N {{{\dot a}_i}(X){Y^{i + 1}} + } \sum\limits_{i = 0}^N {i{a_i}(X){Y^{i - 1}}\frac{1}{H(X)}\left(K_0(X)+K_1(X)Y+\ldots+K_d(X)Y^d\right)}  = \left( {g(X) + h(X)Y} \right)\sum\limits_{i = 0}^N {{a_i}(X){Y^i}}.
\end{equation}
Dot over $a_i(X)$ denotes derivative with respect to $X$.
The degree in variable $Y$ of left hand side (L.H.S) in Eq. \eqref{qred} is $i+d-1$ and the degree in variable $Y$ of right hand side (R.H.S) in Eq. \eqref{qred} is $i+1$. Balancing degrees between both sides we get $i+d-1=i+1$, hence $d=2$. So the method is applicable when the degree of Eq. \eqref{sysode2} in variable $Y$  is less than or equal to 2. Taking the maximum degree 2, Eq. \eqref{qred} can be rewritten as
\begin{equation}\label{qred2}
	\sum\limits_{i = 0}^N {{{\dot a}_i}(X){Y^{i + 1}} + } \sum\limits_{i = 0}^N {i{a_i}(X){Y^{i - 1}}\frac{1}{H(X)}\left(K_0(X)+K_1(X)Y+K_2(X)Y^2\right)}  = \left( {g(X) + h(X)Y} \right)\sum\limits_{i = 0}^N {{a_i}(X){Y^i}}.
\end{equation}

\textbf{Step FIM4}(Derive the Algebraic System of equations for coefficients of $Y^i$): 
Comparing coefficients of $Y^i\;(i=N+1,N,\ldots,1,0)$ on both sides of \eqref{qred2}, and for $H(X)\neq 0$ canceling $H(X)$ in denominator from both sides we obtain
\begin{subequations}\label{Yeq}
	\begin{align}
		\label{Yeq1}&{Y^{N + 1}}:\,\,H(X){{\dot a}_N}(X)+Na_N(X)K_2(X)  = H(X)h(X){a_N}(X),\\
		\label{Yeq2_1}&{Y^N}\,\,\,\,:\,H(X){{\dot a}_{N - 1}}(X) + NK_1(X){a_N}(X) + (N + 1)K_0(X){a_{N + 1}}(X) = H(X)g(X){a_N}(X) + H(X)h(X){a_{N - 1}}(X),\\
		&\,\,\,\,\,\,\,\,\,\,\,\,\,\,\,\,\,\,\,\,\,\,\,\,\,\,\,\,\,\,\,\,\,\,\,\,\,\,\,\,\,\,\,\,\,\,\,\,\,\,\,\,\,\,\,\,\,\,\,\,\, \vdots\nonumber\\
		\label{Yeq3}&{Y^1}\,\,\,\,\,:\,H(X){{\dot a}_0}(X) + K_1(X){a_1}(X) + 2K_0(X){a_2}(X) = H(X)g(X){a_1}(X) + H(X)h(X){a_0}(X),\\
		\label{Yeq4}&{Y^0}\,\,\,\,:\,K_0(X){a_1}(X) = H(X)g(X){a_0}(X), 
	\end{align}
\end{subequations}
where $a_i(X)=0$ for $i<0$ and $i>N$.

In the next step FIM5 we take $a_N=1$ to derive the polynomial forms of $h(X),g(X),a_i(i=0..\ldots N-1)$. For $a_N=1$ from Eq. \eqref{Yeq1} we obtain 
\begin{equation}\label{hform}
	h(X)=\frac{NK_2(X)}{H(X)}.
\end{equation}
If $H (X)$ is not a constant and at the same time degree of $P(X(\xi),Y(\xi))$ in variable $Y$ is 2 then it is clear from Eqs. \eqref{Yeq1}, \eqref{hform} that $h (X) $ will not be polynomial in $X $. In this case we avoid such non-polynomial form of $h (X)$ by making the transformation \cite{mirza}
\begin{equation}\label{trans}
	d\xi=H(X)d\eta,
\end{equation}
in Eq. \eqref{sysode} temporarily. Applying the transformation in Eq. \eqref{sysode} we get
\begin{subequations}\label{sysodetrans}
	\begin{align}
		\label{sysodetrans1}&X_{\eta}(\eta)= H(X)Y,\\
		\label{sysodetrans2}&Y_{\eta}(\eta)=K_0(X)+K_1(X)Y+\ldots+K_d(X)Y^d.
	\end{align}
\end{subequations}

Consequently the nonlinear algebraic system becomes
\begin{subequations}\label{Yeq2}
	\begin{align}
		\label{Yeq12}&{Y^{N + 1}}:\,\,H(X){{\dot a}_N}(X)+Na_N(X)K_2(X)  = h(X){a_N}(X),\\
		\label{Yeq22}&{Y^N}\,\,\,\,:\,H(X){{\dot a}_{N - 1}}(X) + NK_1(X){a_N}(X) + (N + 1)K_0(X){a_{N + 1}}(X) = g(X){a_N}(X) + h(X){a_{N - 1}}(X),\\
		&\;\;\;\;\;\;\;\;\;\;\;\;\;\;\;\;\;\;\;\;\;\;\;\;\;\;\;\;\;\;\;\;\;\;\;\;\vdots\;\;\;\;\;\;\;\;\;\;\;\;\;\;\;\;\;\;\;\;\;\;\;\;\;\;\;\;\;\;\;\;\;\;\;\;\;\;\;\;\;\;\;\;\;\;\;\;\;\;\;\;\vdots\;\;\;\;\;\;\;\;\;\;\;\;\;\;\;\;\;\;\;\;\;\;\;\;\;\;\;\;\nonumber\\
		\label{Yeq32}&{Y^1}\,\,\,\,\,:\,H(X){{\dot a}_0}(X) + K_1(X){a_1}(X) + 2K_0(X){a_2}(X) = g(X){a_1}(X) + h(X){a_0}(X),\\
		\label{Yeq42}&{Y^0}\,\,\,\,:\,K_0(X){a_1}(X) = g(X){a_0}(X). 
	\end{align}
\end{subequations}
Now for $a_N=1$, from Eq. \eqref{Yeq12} we get $h(X)=NK_2(X)$ which is polynomial in $X$. In the following steps, we explain all the procedures with the help of Eq. \eqref{Yeq}, because the same procedures are applicable when Eq. \eqref{Yeq2} is considered for the non-polynomial case of $h(X)$.

\textbf{Step FIM5}(Determine degrees of $h (X), g (X), a_i (X)(i=0\ldots N)$ and express them in polynomial forms): For simplicity substitute $a_N=1$ in Eq. \eqref{Yeq1} (in Eq. \eqref{Yeq12} for non-polynomial case of $h(X)$) and obtain polynomial form of $h(X)$.

To determine polynomial forms of $g,a_i(i<N)$, the degrees of $X$ between L.H.S and R.H.S in each equation of Eqs. \eqref{Yeq} (in Eqs. \eqref{Yeq2} for non-polynomial case of $h(X)$) are balanced. The balancing process is implemented in the following order:
\begin{itemize}
	\item[i.] Replace $g\rightarrow X^{d_g},h(X),a_i\rightarrow X^{d_{a{_i}}}(i<N)$ in each equation from \eqref{Yeq2_1} to \eqref{Yeq4}. Here should be noted, if degree of $P(X(\xi),Y(\xi))$ in variable $Y$ is 2 and $H(X)$ is not a constant (i.e. the non-polynomial case of $h(X)$) then the replacements are done in Eqs. \eqref{Yeq2} instead of Eqs. \eqref{Yeq}.  
	\item[ii.]$d_g,d_{a_{i}}$ are replaced by the sequence $0,1, 2,\ldots$ of all positive integers arranged in increasing order. To avoid an infinite loop, sequences are taken up to a certain maximum number. In our library, the maximum number is 5. Sometimes it is impossible to balance the equations for any positive integer numbers 
	\item[iii.] Take the numbers from sequences at which L.H.S and R.H.S of each equation are balanced in the degree of $X$.
\end{itemize}
After balancing, if $d_{g}, d_{a_{i}}\;(i<N)$ are degrees of $g,a_{i}\;(i<N)$ respectively, then $g,a_{i}$ are expressed by
\begin{subequations}\label{parapoly}
	\begin{align}
		\label{gpoly}&g=g_0+g_1X+\ldots+g_{d_g}X^{d_g},\\
		\label{a0poly}&a_{0}=a_{00}+a_{01}X+\ldots+a_{0d_{a_{0}}}X^{d_{a_{0}}},\\
		\label{a1poly}&a_{1}=a_{10}+a_{11}X+\ldots+a_{1d_{a_{1}}}X^{d_{a_{1}}},\\
		&\vdots=\;\;\;\;\;\;\;\;\;\;\;\;\;\;\vdots\nonumber\\
		\label{aNminus1poly}&a_{N-1}=a_{(N-1)0}+a_{(N-1)1}X+\ldots+a_{(N-1)d_{a_{(N-1)}}}X^{d_{a_{(N-1)}}}.
	\end{align}
\end{subequations}
Where $g_i(i=0,1,\ldots,d_g),a_{ij}(i=0,1,\ldots,N-1,j=0,1,\ldots,d_{a_{i}})$ are arbitrary constants.
In GiNaCDE, all possible combinations of balanced degrees (whose values are $<$ 6) for $g,a_i$ are calculated, and for each combination, the solutions of input-NLPDE are derived.\\

\textbf{Step FIM6}(Derive the Nonlinear Algebraic System for the parameters $g_i,a_{ij}$): Substitute $a_N=1$ and \eqref{parapoly} into each equation of \eqref{Yeq}. The coefficients of the power in $X,Y$ in each equation of \eqref{Yeq} must vanish. Collect the coefficients and generate a nonlinear algebraic system of equations parametrized by $g_i,a_{ij},\boldsymbol{K},\boldsymbol{P}$, integrating constants (for integrable NLPDE) and parameters appearing in input-NLPDE.\\

\textbf{Step FIM7}(Solve the Nonlinear Parameterized Algebraic System): Here, the external parameters are parameters appearing in input-NLPDE and integrating constants. Internal parameters are $g_i,a_{ij},\boldsymbol{K},\boldsymbol{P}$. The nonlinear algebraic system is solved following a similar process in step F4. Like step F4, here also, the runtime of this algorithm mainly depends on this step.\\

\textbf{Step FIM8}(Build solutions): The solutions in step FIM7 are substituted in Eq. \eqref{irred} and using $Y(\xi)=U_\xi(\xi)$, Eq. \eqref{irred} converts into first-order NLODE called first integral form of Eq. \eqref{tweq2}. Some well-known forms of first-order NLODE with solutions have been listed in Appendix: \ref{sec:appenA}, \ref{sec:appenB}, \ref{sec:appenC}, \ref{sec:appenD}. If first integral form matches with any form of listed NLODE in Appendix, solutions are shown, otherwise the program shows only the first integral forms. Combining Eqs. \eqref{twtrans}, \eqref{twco}, we obtain final solutions of input-NLPDE in original variables.\\

\subsection{Explanation of each step with a working example}
To illustrate every step of the above algorithm for the first integral method, like the F-expansion method, we have also used the NLS equation \eqref{nls}. 

\textbf{Step FIM1:} Similar to F-expansion method Eq. \eqref{nls} is converted to Eq. \eqref{nlsNLODE} with the condition \eqref{nlscond}.

\textbf{Step FIM2:} Letting $U=X$, we recast Eq. \eqref{nlsNLODE} as a first-order system in traveling-wave coordinate $(\xi)$. So,
\begin{align}\label{nls1storder}
	&X_\xi=Y,\nonumber\\
	&Y_\xi=\frac{1}{{p{k_1}^2}}\left( {\left( { - {p_0} +  \frac{{{k_0}^2}}{{4p{k_1}^2}}} \right)X + q{X^3}} \right).
\end{align}
Here $H(X)=p{k_1}^2$.

\textbf{Step FIM3:} We set $N=1$, and from Eq. \eqref{irred} we get
\begin{equation}\label{nlsirred}
	a_0(X)+a_1(X)Y=0,
\end{equation}
and from Eq. \eqref{qred2} we get
\begin{equation}\label{nlsqred}
	\sum\limits_{i = 0}^1 {{{\dot a}_i}(X){Y^{i + 1}} + } \sum\limits_{i = 0}^1 {i{a_i}(X){Y^{i - 1}}\frac{1}{{p{k_1}^2}}\left( {\left( { - {p_0} +  \frac{{{k_0}^2}}{{4p{k_1}^2}}} \right)X + q{X^3}} \right)}  = \left( {g(X) + h(X)Y} \right)\sum\limits_{i = 0}^1 {{a_i}(X){Y^i}},
\end{equation}

\textbf{Step FIM4:} After substitution of \eqref{nls1storder} into \eqref{Yeq} and taking $N=1$, one gets

\begin{subequations}\label{nlsYeq}
	\begin{align}
		\label{nlsYeq1}&4{p^2}k_1^4{{\dot a}_1}(X) = 4{p^2}k_1^4{a_1}(X)h(X),\\
		\label{nlsYeq2}&4{p^2}k_1^4{{\dot a}_0}(X) = 4gk_1^4{p^2}{a_1}(X) + 4k_1^4{p^2}{a_0}(X)h(X), \\
		\label{nlsYeq3}&4pqk_1^2{a_1}(X){X^3} - 4p{p_0}k_1^2{a_1}(X)X + k_0^2{a_1}(X)X = 4{p^2}k_1^4{a_0}(X)g(X).
	\end{align}
\end{subequations}
Taking $a_1(X)=1$, from Eq. \eqref{nlsYeq1} we deduce $h(X)=0$ that is polynomial. So the transformation \eqref{trans} is not required here and we carry out our next derivations with Eqs. \eqref{nlsYeq}.

\textbf{Step FIM5:} $h(X)=0$ was derived in previous step. Now replacing $g\rightarrow X^{d_g},h(X)\rightarrow 0,a_0\rightarrow X^{d_{a0}}$ into Eqs. \eqref{nlsYeq2}, \eqref{nlsYeq3}, the exponents of $X$ of L.H.S and R.H.S are collected by two separate list {\em lhs} and {\em rhs} respectively. After collections, {\em lhs, rhs} are
\begin{align}
	&lhs[\text{Eq. 43b}]=\{d_{a0}-1\},\;rhs[\text{Eq. 43b}]=\{1\};\\
	&lhs[\text{Eq. 43c}]=\{3,1\},\;rhs[\text{Eq. 43c}]=\{d_{a0}+d_g\};
\end{align}
Then the values of $d_g,d_{a0}$ are increased from $0$ to $5$. At $d_g=1,d_{a0}=2$, we get $max(lhs[\text{Eq. 43b}])=max(rhs[\text{Eq. 43b}])$ and $max(lhs[\text{Eq. 43c}])=max(rhs[\text{Eq. 43c}])$ which detects balancing is successful, and using the balancing degrees $d_g=1,d_{a0}=2$, $g,a_0$ are expressed by
\begin{align}\label{nlsparapoly}
	g = g_0+g_1X,
	a_0 = a_{00}+a_{01}X+a_{02}X^2,
\end{align}
where $g_0,g_1,a_{00},a_{01},a_{02}$ are arbitrary constants.

\textbf{Step FIM6:} Substituting $a_1=1$ and Eq. \eqref{parapoly} into each equation of Eqs. \eqref{nlsYeq} and collecting coefficients of $X^iY^j(i=0,1,2,3,j=0,1)$, we obtain the following set of nonlinear algebraic equations:
\begin{subequations}\label{nlsalgset}
	\begin{align}
		&X^0Y^0: -4g_0k_1^4p^2a_{00} = 0,\\
		&X^1Y^0: -4g_0k_1^4a_{01}p^2-4k_1^4p^2a_{00}g_1-4p_0k_1^2p+k_0^2 = 0,\\
		&X^2Y^0: -4g_0a_{02}k_1^4p^2-4k_1^4a_{01}p^2g_1 = 0,\\
		&X^3Y^0: 4k_1^2pq-4a_02k_1^4p^2g_1 = 0,\\
		&X^0Y^1: 4k_1^4a_{01}p^2-4g_0k_1^4p^2 = 0,\\
		&X^1Y^1: 8a_{02}k_1^4p^2-4k_1^4p^2g_1 = 0.
	\end{align}
\end{subequations}

\textbf{Step FIM7:} Here, the internal parameters are $g_0,g_1,a_{00},a_{01},a_{02},k_0,k_1,p_0,p_1$ and external parameters are $p,q$. Solving the set of equations \eqref{nlsalgset} for the internal parameters only, the solutions are
\begin{subequations}
	\begin{align}
		\label{nlsalgsetSolu1}&{a_{01}} = 0,{a_{02}} = \frac{{{g_1}}}{2},{g_0} = 0,{k_1} =  \pm \frac{{\sqrt {2q} }}{{\sqrt p {g_1}}},{p_0} =  \frac{{{k_0}^2{g_1}^3 - 16 {a_{00}} {q^2}}}{{8q{g_1}}};\\
		\label{nlsalgsetSolu2}&{a_{00}} = 0,{a_{01}} = 0,{a_{02}} = \frac{{{g_1}}}{2},{g_0} = 0,{k_1} =  \pm \frac{{\sqrt {2q} }}{{\sqrt p {g_1}}},{p_0} =  \frac{{{k_0}^2{g_1}^2}}{{8q}}.
	\end{align}
\end{subequations}

\textbf{Step FIM8:} Substituting \eqref{nlsalgsetSolu1} into \eqref{nlsirred} the first integral form to Eq. \eqref{nlsNLODE} is 
\begin{equation}\label{nls_fif1}
	{U_\xi } + {a_{00}} + \frac{{{g_1}}}{2} {U^2} = 0.
\end{equation}
Equation \eqref{nls_fif1} is Riccati equation and to find its solutions Appendix: \ref{sec:appenA} is used. Finally, using \eqref{nlscond} and substituting \eqref{nlsalgsetSolu1}, $5$ exact solutions of Eq. \eqref{nls} can be obtained. One can note that when $g_1=\sqrt{\frac{-2q}{p}}$, we have the same first integral form as given in \cite{fim7}.
 
\section{Algorithm Implementation}\label{sec:imple}
We have implemented the proposed algorithms described in Secs. \ref{sec:Fexpn}, \ref{sec:mF} and \ref{sec:fim}, into GiNaCDE. GiNaCDE is a C++ library that is built on a pure C++ symbolic library GiNaC \cite{ginac}. Besides this library version, we have also developed a GUI version of GiNaCDE called GiNaCDE GUI. When we solve differential equations using GiNaCDE GUI, we do not have to write any C++ code, and compilation of any code is not required. This GUI version guides us in each step to obtain the output results. However, a complete guide of GiNaCDE GUI has been provided with the GiNaCDE source code. In both versions of GiNaCDE (library and GUI), the output results are saved in a text file with calculating steps. Output results can be saved in Maple or Mathematica programming language by assigning the C++ macros {\em Maple} or {\em Mathematica} to the programming variable {\em {output}}. 

In order to start a solution process for a given NLPDE or NLODE in GiNaCDE, we require some initial data (the options and parameters specified by the user). Some programming variables such as \textit{twcPhase, positivePart, negativePart, NValue,  degAcoeff, ASolve, paraInDiffSolve} (detailed descriptions of these variables are given in the software user manual) are available, which can be initially set up by the users in their own choices for getting better results for a given NLPDE before starting the solution process in the GiNaCDE software. 
In all three methods, if the input NLPDE or NLODE  is complex, then the software tries to separate the real and imaginary parts following the step F1.
In all three methods, if the input NLPDE or NLODE is integrable, the software tries to integrate them after starting the solution process. If the integration is successful, then the software gives an option to us to assign a numerical or symbolic value to the integrating constant(s) $ic_i\;(i=1,2,..)$ in our choices. All these options make the software more powerful and flexible, enhancing its ability to find many new exact solutions to huge variants of NLPDEs.

Now we shall discuss some implementation details of each method separately:

\textbf{F-expansion and mF-expansion methods:} These methods have been implemented in F\_expns\_methd.cpp and F\_expns\_methd.h files.
The F-expansion and mF-expansion methods are chosen by the C++ macros {\em F\_expansion, mF\_expansion} respectively. 
One can use the coordinates $\boldsymbol{K},\boldsymbol{P}$ in own choices with the help of the programming variable \textit{twcPhase}.
We can take any one or both parts in the solutions \eqref{soluseries} with the help of the programming variables \textit{positivePart}, \textit{negativePart}. In GiNaCDE, we set the maximum allowed value of $N$ at $10$. However, sometimes in some cases, the auto-evaluated value of $N$ exceeds 10. Then this step fails to find $N$.
In this case, we can check the solutions of input-NLPDE by specifying the value of $N$ in our choice lower than $10$ with the help of the programming variable {\em NValue}. If we do not assign any value to the variable {\em NValue}, the value of $N$ is auto-evaluated following the criteria in step F2.
In these methods, we initially have to input A.E. to start the solution process in GiNaCDE manually. One can choose the parameters $A_i(i=0,1,\ldots \delta)$ in own choices with the help of the programming variable \textit{degAcoeff}. 
The variable {\em ASolve} confirms whether the nonlinear algebraic system will be solved for the parameters contained in A.E. ( i.e., the parameters $A_i$ where $i=0,1,\ldots \delta$) along with other parameters. The parameters appearing in input-NLPDE (these parameters are belong to external parameters) are supplied in the programming variable {\em paraInDiffSolve} to solve the nonlinear algebraic system for those parameters also.
This will determine the conditions on that external parameters so that exact solutions are obtained.

\textbf{FIM:} This method has been implemented in fim.cpp and fim.h files. This method is chosen by the C++ macro {\em FIM}. For the first integral method, we do not have to input A.E. Here, for initializations, we have only three programming variables \textit{twcPhase, NValue, paraInDiffSolve}. The variables \textit{twcPhase, paraInDiffSolve} have been discussed above. Like F-expansion method the value of $N$ is also assigned by the variable {\em NValue} and the default value is $N=1$. However, in our library, the allowed values of {\em NValue} are 1 and 2.


\section{Application}\label{sec:examp}
In this paper, we have presented the algorithms that automate three different methods which can be applied to determine exact solutions of the wide variant of NLPDEs. F-expansion and mF-expansion methods can be applied to higher-order NLPDEs. But, FIM method is applicable to an NLPDE when its transformed NLODE \eqref{tweq2} is second-order only. Actually, there are no rules to know in advance the appropriate method among three to solve the given NLPDE. To illustrate the effectiveness of the algorithms, in the following subsections, using the GiNaCDE library, we have applied all these three methods to solve some NLPDEs, including higher-order and complex NLPDEs.

To start a solution process for a given NLPDE, we require some initializations (the options and parameters specified by the user). There is no rule to know in advance a specific type of initialization to get better results. However, we have searched some initializations for solving the following five examples so that the manually chosen initializations give better results. In the following examples, such initializations for each NLPDE are also described. The chosen assigned values of the integrating constant(s) of the integrable NLPDEs among the following examples are also described. 
In the case of F-expansion and mF-expansion methods, there are no rules to know in advance what type of A.E. can give exact solutions for a given NLPDE. In this context, we have shown the solutions of some well-known A.E. in the Appendix: \ref{sec:appenA}, \ref{sec:appenB}, \ref{sec:appenC}, \ref{sec:appenD}.
We have checked the solutions for the NLPDEs listed below using some specific types of A.E. These specific forms of A.E. and the corresponding number of solutions with calculating times are shown in Table \ref{compareTab}.
Here one can note that more new types of solutions of the NLPDEs can be obtained using different forms of A.E.

We have well described the procedure to get the exact solutions of an NLPDE using GiNaCDE in the user manual. 
We have solved the following examples using the C++ code as given in the examples folder provided with the GiNaCDE source code.
However, in this paper, we do not quote the algebraic expressions of the exact solutions obtained in the following examples since their final expressions are too lengthy. We have observed that some of the solutions calculated by us were previously reported in the literature.
Interestingly in some cases, new solutions are also obtained, and some new solutions are shown here.

\begin{table}[H]
	\centering
	\caption{Number of solutions of some NLPDEs with different types of A.E. Solutions are computed with GiNaCDE library. Test runs performed on a Dell laptop with Intel(R) Core(TM) i5-7200U CPU 2.50 GHz, 8 GB of RAM.}
	\begin{tabular}{|c|c|c|c|c|c|c|c|c|}
		\hline
		\multirow{2}{*}{NLPDEs}& \multicolumn{3}{c|}{F-expansion} &  \multicolumn{3}{c|}{mF-expansion}  &  \multicolumn{2}{c|}{FIM}  \\
		\cline{2-9}
		& A.E. &\#Sols. &$\begin{matrix}
			\text{CPU}\\
			\text{time (s)}
		\end{matrix}$&  A.E. &\#Sols. &$\begin{matrix}
			\text{CPU}\\
			\text{time (s)}
		\end{matrix}$& \#Sols. &$\begin{matrix}
			\text{CPU}\\
			\text{time (s)}
		\end{matrix}$  \\
		\hline
		Eq. \eqref{eckhaus}&  $F^\prime=\sqrt{A_0+A_1F}$&  0&  0.4&  $F^\prime={A_1F+A_2F^2}$&  5&  0.9&  5& 1.2 \\
		\hline
		Eq. \eqref{sSK}&  $F^\prime=\sqrt{A_0+A_2F^2+A_4F^4}$&  40&  6.5&  $F^\prime={F+F^2}$&  4&  13.8&  -& - \\
		\hline
		Eq. \eqref{gKdV}&  $F^\prime=\sqrt{A_1F+A_2F^2+A_3F^3}$&  10&  2.4&  $F^\prime={1+F+F^2}$&  25&  50&  -&-  \\
		\hline
		Eq. \eqref{nlseWtKerr}&  $F^\prime=\sqrt{A_2F^2+A_3F^3+A_4F^4}$&  9&  1.3&  $F^\prime={A_0+A_1F+A_2F^2}$&  0&  33.8&  0&1.4  \\
		\hline
		Eq. \eqref{Ks}&  $F^\prime=\sqrt{A_0+A_1F+A_2F^2+A_3F^3}$&  0&  0.56&  $F^\prime={A_0+A_1F+A_2F^2}$&  14&  34&  4&1.7  \\
		\hline
	\end{tabular}
	\label{compareTab}
\end{table}

\subsection{Eckhaus equation}
Let us consider the Eckhaus equation \cite{eckhaus}
\begin{equation}\label{eckhaus}
	I{u_t} + {u_{xx}} + 2{\left( {{{\left| u \right|}^2}} \right)_x}u + {\left| u \right|^4}u = 0,
\end{equation}
which belongs to a different class of integrable NLPDEs, often termed C-integrable equations, i.e. linearizable through a change of dependent variables, into the linear, free Schr\"{o}dinger equation \cite{eckhaus2}. In seeking exact solutions of Eq. \eqref{eckhaus}, many powerful methods have been applied, such as the first-integral method \cite{eckhaus3}, ${G'}/G$-expansion method \cite{eckhaus4}.

The initializations we have used to solve the Eq. \eqref{eckhaus} by employing each method available in the GiNaCDE library are given below:
\begin{center}
	\begin{tabular}{|c|c|c|}
		\hline
		\textbf{F-expansion:}&  \textbf{mF-expansion:}& \textbf{FIM:} \\
		\hline
		\begin{minipage}{2.3in}
			\small
			\begin{verbatim}
				twcPhase=
				lst{lst{k_0,k_1},lst{p_0,p_1}};
				degAcoeff=lst{1,A_0,A_1};
				ASolve=false;
				positivePart=true; 
				negativePart=true;
				paraInDiffSolve=lst{};
				filename="Eckhaus_Fexp.txt";
			\end{verbatim}
		\end{minipage}&  	
		\begin{minipage}{2.3in}
			\small
			\begin{verbatim}
				twcPhase=
				lst{lst{k_0,k_1},lst{p_0,p_1}};
				degAcoeff=lst{2,0,A_1,A_2};
				ASolve=false;
				positivePart=true; 
				negativePart=true;
				paraInDiffSolve=lst{};
				filename="Eckhaus_mF.txt";
			\end{verbatim}
		\end{minipage}&  
		\begin{minipage}{2.3in}
			\small
			\begin{verbatim}
				twcPhase=
				lst{lst{k_0,k_1},lst{p_0,p_1}};
				paraInDiffSolve=lst{};
				filename="Eckhaus_FIM.txt";
			\end{verbatim}
		\end{minipage}\\
		\hline
	\end{tabular}
\end{center}

In the all methods (F-expansion, mF-expansion and FIM), after substituting the traveling-wave solution \eqref{nls_twtrans} into Eq. \eqref{eckhaus}, it is transformed into NLODE. Then this NLODE is separated into real and imaginary parts. The imaginary part is integrable and after one time integration GiNaCDE itself assigns the integrating constant to $0$. Then from imaginary part, following the step F1 in the algorithm of automated F-expansion method, GiNaCDE derives the condition
\begin{equation}\label{eckhausCond}
	k_1 = -\frac{k_0}{2p_1}.
\end{equation}
In F-expansion and mF-expansion methods, value of $N$ is $\frac{1}{2}$ which is auto-evaluated following the step F2 in our proposed algorithm.
With the condition \eqref{eckhausCond}, solutions of the real part are evaluated in all three methods. In F-expansion method, we do not get any solution. In mF-expansion method, using the solutions of Bernouli equation given in \eqref{bernouli_solu} with condition \eqref{eckhausCond}, the solutions to Eq. \eqref{eckhaus} are obtained.

\subsection{Seventh-order Sawada–Kotara equations}
J. Feng \cite{sSK} considered the following Seventh-order Sawada-Kotera equation (sSK)
\begin{equation}\label{sSK}
	{u_t} + {(63{u^4} + 63(2{u^2}{u_{2x}} + uu_x^2) + 21(u{u_{4x}} + u_{2x}^2 + {u_x}{u_{3x}}) + {u_{6x}})_x} = 0,
\end{equation}
to find out some closed-form trigonometric, hyperbolic, rational solutions applying $G'/G$ expansion method. Several methods such as  Adomian decomposition method (ADM) \cite{sSK1}, He’s variational iteration method \cite{sSK2}, Reconstruction of Variational Iteration Method (RVIM) \cite{sSK3}, has been applied for computing approximated solutions to Eq. \eqref{sSK}. 

The initializations we have used to solve the Eq. \eqref{sSK} by employing each method available in the GiNaCDE library are given below:
\begin{center}
	\begin{tabular}{|c|c|c|}
		\hline
		\textbf{F-expansion:}&  \textbf{mF-expansion:}& \textbf{FIM:} \\
		\hline
		\begin{minipage}{2.3in}
			\small
			\begin{verbatim}
				twcPhase=lst{lst{k_0,k_1},lst{}};
				degAcoeff=lst{4,A_0,0,A_2,0,A_4};
				ASolve=true;
				positivePart=true; 
				negativePart=false;
				paraInDiffSolve=lst{};
				filename="7thorder_Fexp.txt";
			\end{verbatim}
		\end{minipage}&  	
		\begin{minipage}{2.3in}
			\small
			\begin{verbatim}
				twcPhase=lst{lst{k_0,k_1},lst{}};
				degAcoeff=lst{2,0,1,1};
				ASolve=false;
				positivePart=true; 
				negativePart=false;
				paraInDiffSolve=lst{};
				filename="7thorder_mF.txt";
			\end{verbatim}
		\end{minipage}&  
		\begin{minipage}{2.3in}
			\small
			\begin{verbatim}
				twcPhase=lst{lst{k_0,k_1},lst{}};
				paraInDiffSolve=lst{};
				filename="7thorder_FIM.txt";
			\end{verbatim}
		\end{minipage}\\
		\hline
	\end{tabular}
\end{center}
In the all methods, after transforming Eq. \eqref{sSK} into NLODE using the traveling-wave coordinate $\xi$ given by
\begin{equation}\label{KdVB_twtrans}
	u(t,x)=U(\xi)\;\;\text{where}\;\;\xi=k_0t+k_1x,
\end{equation}
GiNaCDE integrates this NLODE one time, and we have assigned the integrating constant $ic_1$ to $0$ in our choice. In FIM, no solution is obtained. In the case of the F-expansion method, the library returns some new Jacobian solutions using the known solutions of A.E. given in \eqref{024_solu}. Total 40 new solutions are obtained. For simplicity only $3$ new solutions are presented here: 
\begin{align}
	&u(t,x)=  {a_0} + \frac{{a_2^2 \left( { - {A_2}^2 + \sqrt {{A_2}^2S} } \right)}}{{2{A_2} {A_4}}}{\text{JacobiS}}{{\text{N}}^2}\left( {\frac{{\sqrt {2{A_2} \left( { - {A_2}^2 - \sqrt {{A_2}^2S} } \right)} \xi }}{{2{A_2}}},\frac{{\sqrt {2{A_0} {A_4} \left( { - 2 {A_0} {A_4} + {A_2}^2 - \sqrt {{A_2}^2S} } \right)} }}{{2{A_0} {A_4}}}} \right), \hfill \\
	&u(t,x)=   {a_0} +  \frac{{a_2^2 \left( { - {A_2}^2 + \sqrt {{A_2}^2S} } \right)}}{{2{A_2} {A_4}}}{\text{JacobiC}}{{\text{N}}^2}\left( {\frac{{\sqrt { - {A_2} \sqrt {{A_2}^2S} } \xi }}{{{A_2}}},\frac{{\sqrt { - 2S\left( {\sqrt {{A_2}^2S}  - S} \right)} }}{{2S}}} \right), \hfill \\
	&u(t,x)=   {a_0} +  \frac{{a_2^2 \left( { - {A_2}^2 - \sqrt {{A_2}^2S} } \right)}}{{2{A_2} {A_4}}}{\text{JacobiD}}{{\text{N}}^2}\left( { \frac{{\sqrt {2{A_2} \left( {{A_2}^2 + \sqrt {{A_2}^2S} } \right)} \xi }}{{2{A_2}}}, \frac{{\sqrt {2{A_0} {A_4} \left( {\sqrt {{A_2}^2S}  - S} \right)} }}{{2{A_0} {A_4}}}} \right), 
\end{align}  
where
\begin{align}
	&  {A_4} =  - \frac{{{a_2}}}{{2k_1^2}}, {a_2} =  -  \frac{{48 {A_2} {a_0} {k_1}^2 + 63 {a_0}^2 + 8 {A_2}^2{k_1}^4 \pm 2 \sqrt 2 l}}{{6{A_0} {k_1}^2}}, \hfill \\
	&  \xi  = \left( { - 4 \left( {24 {A_2}^3{k_1}^6 + 132 {A_2}^2{a_0} {k_1}^4 + 63 {a_0}^3 + \left( {189 {a_0}^2{k_1}^2 \pm 2 \sqrt 2 {k_1}^2l} \right){A_2}} \right){k_1}} \right)t + {k_1}x, \hfill \\
	&  l = \sqrt {\left( {4 {A_2}^2{k_1}^4 + 42 {A_2} {a_0} {k_1}^2 + 63 {a_0}^2} \right)\left( {{A_2} {k_1}^2 + 3 {a_0}} \right)\left( {2 {A_2} {k_1}^2 + 3 {a_0}} \right)}, S = \left( { - 4 {A_0} {A_4} + {A_2}^2} \right).
\end{align}
In the above solutions two sign combinations in each solution are to be taken.

\subsection{Fifth-order Generalized KdV equation}
Consider the general fifth-order KdV equation (gKdV)\cite{gKdV,gKdV1,gKdV2,gKdV3,gKdV4}
\begin{equation}\label{gKdV}
	pu{u_{3x}} + q{u_x}{u_{2x}} + r{u^2}{u_x} + {u_{5x}} + {u_t} = 0,
\end{equation}
where $p,q,r$ are arbitrary non-zero and real parameters. The model equation \eqref{gKdV} has a wide range of applications in many important physical phenomenon including quantum mechanics, nonlinear optics, plasma physics, one dimensional nonlinear lattice. The values of the parameters $p,q,r$ will drastically change the characteristics of the gKdV equation \eqref{gKdV}. Many well-known equations  can be constructed from the gKdV equation by changing these parameters, such as: the Lax equation $\left(q=2p,r=\frac{3p^2}{10}\right)$, the Sawada–Kotera (SK) equation $\left(q=p,r=\frac{p^2}{5}\right)$, the generalized Kaup–Kupershmidt equation (GKK)$\left(q=\frac{5p}{2},r=\frac{p^2}{5}\right)$, the generalized Ito equation (GI) $\left(q=2p,r=\frac{2p^2}{9}\right)$. A variety of powerful and direct methods have been applied to obtain exact solutions to Eq. \eqref{gKdV}. Among of them are: the extended tanh method \cite{gKdV}, the generalized tanh-coth method \cite{gKdV1}, an extended Jacobian elliptic function expansion approach \cite{gKdV2}, Exp-Function method \cite{gKdV3,gKdV4}. Zhi-bin Li et al. in their works \cite{rath} have found out some new conditions among $p,q,r$ for which some new solutions were obtained using their computer package RATH. We ask GiNaCDE to derive solutions of Eq. \eqref{gKdV} by applying all three methods available in the library. 

We have used the following initializations:
\begin{center}
	\begin{tabular}{|c|c|c|}
		\hline
		\textbf{F-expansion:}&  \textbf{mF-expansion:}& \textbf{FIM:} \\
		\hline
		\begin{minipage}{2.3in}
			\small
			\begin{verbatim}
				twcPhase=lst{lst{k_0,k_1},lst{}};
				degAcoeff=lst{3,0,A_1,A_2,A_3};
				ASolve=false;
				positivePart=true; 
				negativePart=false;
				paraInDiffSolve=lst{r};
				filename="5thGKdV_Fexp.txt";
			\end{verbatim}
		\end{minipage}&  	
		\begin{minipage}{2.3in}
			\small
			\begin{verbatim}
				twcPhase=lst{lst{k_0,k_1},lst{}};
				degAcoeff=lst{2,1,1,1};
				ASolve=false;
				positivePart=true; 
				negativePart=false;
				paraInDiffSolve=lst{q};
				filename="5thGKdV_mF.txt";
			\end{verbatim}
		\end{minipage}&  
		\begin{minipage}{2.3in}
			\small
			\begin{verbatim}
				twcPhase=lst{lst{k_0,k_1},lst{}};
				paraInDiffSolve=lst{};
				filename="5thGKdV_FIM.txt";
			\end{verbatim}
		\end{minipage}\\
		\hline
	\end{tabular}
\end{center}

In the all methods, GiNaCDE transforms Eq. \eqref{gKdV} into traveling-wave coordinate $\xi$ using the Eq. \eqref{KdVB_twtrans} and integrates it one time. We assign the numerical value $0$ to integrating constant $ic_1$.  Interestingly, in the case of F-expansion method, it finds out some new exact solutions with a new condition
\begin{equation}\label{rCond}
	r =  -  \frac{{3\left( {3 p - q} \right)\left( {p - q} \right)}}{8},
\end{equation}
which is completely different from a list of conditions among $p,q,r$ derived in \cite{rath}.
For the condition \eqref{rCond}, with the help of solutions \eqref{123_solu} we obtain $10$ new Jacobian elliptic solutions. For brevity we quote only following $3$ new solutions
\begin{align}
	&u(t,x)=  a_0  + \frac{{6k_1^2{A_3}\left( { - {A_2} + \sqrt S } \right)}}{{2{A_3}(p - q)}}{\text{JacobiS}}{{\text{N}}^2}\left( { \frac{{\sqrt {2{A_1} {A_3}} \xi }}{{2\sqrt { - {A_2} + \sqrt S } }}, \frac{{\sqrt {{A_2}^2 - {A_2} \sqrt S  - 2 {A_1} {A_3}} }}{{\sqrt {2{A_1} {A_3}} }}} \right), \hfill \\
	&u(t,x)=   a_0 -  \frac{{12{A_1}k_1^2{A_3}}}{{(p - q)\left( {{A_2} - \sqrt S } \right)}}{\text{JacobiC}}{{\text{N}}^2}\left( { \frac{{\sqrt[4]{S}\xi }}{2}, \frac{{\sqrt {{A_2} + \sqrt S } }}{{\sqrt 2 \sqrt[4]{S}}}} \right), \hfill \\
	&u(t,x)=   a_0 - \frac{{12{A_1}k_1^2{A_3}}}{{(p - q)\left( {{A_2} - \sqrt S } \right)}}{\text{JacobiD}}{{\text{N}}^2}\left( {\frac{{\sqrt {2{A_1} {A_3}} \xi }}{{2\sqrt {{A_2} - \sqrt S } }}, \frac{{\sqrt {4 {A_1} {A_3} + \sqrt S {A_2} - {A_2}^2} }}{{\sqrt {2{A_1} {A_3}} }}} \right), 
\end{align}
where
\begin{align}
	& a_0 = \frac{{2{A_2}k_1^2}}{{(p - q)}},
	  S =  - 4 {A_1} {A_3} + {A_2}^2,
	 \xi  = \left( { - \frac{{\left( {3 {A_3} {A_1} - {A_2}^2} \right)\left( {3 p - q} \right){k_1}^5}}{{2 p - 2 q}}} \right)t + {k_1}x.
\end{align}

\subsection{Perturbed NLS Equation with Kerr Law Nonlinearity}
We consider the perturbed NLS equation with Kerr law nonlinearity \cite{nlsewtkerr,nlsewtkerr1,nlsewtkerr2,fim4,nlsewtkerr3}
\begin{equation}\label{nlseWtKerr}
	I{u_t} + {u_{2x}} + A|u{|^2}u + I\left( {{G_1}{u_{3x}} + {G_2}|u{|^2}{u_x} + {G_3}{{\left( {|u{|^2}} \right)}_x}u} \right) = 0,
\end{equation}
where $u(t,x)$ represents the complex function and the parameters $G_1,G_2$ and $G_3$ are the higher order dispersion coefficient, the coefficient of Raman scattering, the coefficient of nonlinear dispersion term respectively, while $A$ represents fiber loss. 
The model equation \eqref{nlseWtKerr} has important application in various fields, such as semiconductor materials, optical fiber communications, plasma physics, fluid and solid mechanics. Several methods for finding the exact solutions to \eqref{nlseWtKerr} have been applied, such as first integral method \cite{fim4}, the improved $\tan \left(\frac{\phi(\xi)}{2}\right)$ -expansion method \cite{nlsewtkerr1}, the modified trigonometric function series method \cite{nlsewtkerr2}, the modified mapping method and the extended mapping method \cite{nlsewtkerr3}.
G. Akram et al. \cite{nlsewtkerr} recently have successfully applied the extended $G'/G^2$-expansion method and the first integral method on Eq. \eqref{nlseWtKerr} to find some new exact solutions, which include hyperbolic function solutions, trigonometric function solutions, rational function solutions and soliton solutions.
We run GiNaCDE software applying all the three available methods on Eq. \eqref{nlseWtKerr}. Here we have used the following initializations:

\begin{center}
	\begin{tabular}{|c|c|c|}
		\hline
		\textbf{F-expansion:}&  \textbf{mF-expansion:}& \textbf{FIM:} \\
		\hline
		\begin{minipage}{2.3in}
			\small
			\begin{verbatim}
				twcPhase=
				lst{lst{k_0,k_1},lst{p_0,p_1}};
				degAcoeff=
				lst{4,0,0,A_2,A_3,A_4};
				ASolve=true;
				positivePart=true; 
				negativePart=true;
				paraInDiffSolve=lst{};
				filename="kerrNLS_Fexp.txt";
			\end{verbatim}
		\end{minipage}&  	
		\begin{minipage}{2.3in}
			\small
			\begin{verbatim}
				twcPhase=
				lst{lst{k_0,k_1},lst{p_0,p_1}};
				degAcoeff=lst{2,A_0,A_1,A_2};
				ASolve=false;
				positivePart=true; 
				negativePart=true;
				paraInDiffSolve=lst{};
				filename="kerrNLS_mF.txt";
			\end{verbatim}
		\end{minipage}&  
		\begin{minipage}{2.3in}
			\small
			\begin{verbatim}
				twcPhase=
				lst{lst{k_0,k_1},lst{p_0,p_1}};
				paraInDiffSolve=lst{};
				filename="kerrNLS_FIM.txt";
			\end{verbatim}
		\end{minipage}\\
		\hline
	\end{tabular}
\end{center}

In all methods, the software substitutes the traveling-wave solution \eqref{nls_twtrans} into Eq. \eqref{nlseWtKerr}, and separates the real part and the imaginary part following the step F1. The imaginary part is integrated one time and the software assigns the constant of integration to zero. Then the algebraic expressions of real part and imaginary part are compared and the software detects that they are same equations subject to the following conditions
\begin{equation}\label{nlseWtKerrCond}
	\frac{{{k_0} + 2{p_1}{k_1} - 3{G_1}{k_1}p_1^2}}{{ - p_1^2 - {p_0} + {G_1}p_1^3}} = \frac{{{G_1}k_1^3}}{{ - 3{G_1}{p_1}k_1^2 + k_1^2}} = \frac{{2{k_1}{G_3} + {k_1}{G_2}}}{{3A - 3{p_1}{G_2}}}.
\end{equation}
Same conditions were obtained in \cite{nlsewtkerr3}. Therefore GiNaCDE evaluates the exact analytical solutions of imaginary part only. 

In the case of F-expansion method, using the solutions \eqref{234_solu} GiNaCDE produces two new traveling-wave solutions given by
\begin{align}
	&\mathsmaller{u(t,x)=  {a_0} - \frac{{{a_1}{A_2} {A_3} {{\left( {{\text{sech}}\left( { \frac{{\sqrt {{A_2}} }}{2}\xi } \right)} \right)}^2}}}{{{A_3}^2 - {A_2} {A_4} {{\left( {1 - \tanh \left( {\frac{{\sqrt {{A_2}} }}{2}\xi } \right)} \right)}^2}}} ,\;u(t,x)=  {a_0} + \frac{{2{a_1}{A_2} {\text{sech}}\left( {\sqrt {{A_2}} \xi } \right)}}{{\sqrt { - 4 {A_2} {A_4} + {A_3}^2}  - {A_3} {\text{sech}}\left( {\sqrt {{A_2}} \xi } \right)}}},
\end{align}
where
\begin{align}
	&  \xi  = {k_0}t - \left( { \frac{{3{k_0}}}{l}} \right)x, \;  {A_2} =  - \frac{{2{a_0}^2\left( {2{G_3} +  {G_2}} \right){l}}}{{27 {G_1} {k_0}^2}},\;  {A_3} =  - \frac{{2{a_0}{a_1} \left( {2 {G_3} +  {G_2}} \right) {l}}}{{27 {G_1} {k_0}^2}}, \;  {A_4} =  - \frac{{{a_1}^2\left( {2 {G_3} + {G_2}} \right){l}}}{{54 {G_1} {k_0}^2}},\\
	&l={\left( { - 9 {G_1} {p_1}^2 + {a_0}^2{G_2} + 2 {a_0}^2{G_3} + 6 {p_1}} \right)}^2.
\end{align}

\subsection{Kudryashov–Sinelshchikov Equation}
Now, we study the following Kudryashov–Sinelshchikov equation proposed in \cite{ks,ks1}:
\begin{equation}\label{Ks}
	{u_{3x}} + gu{u_x} - n{u_{2x}} - \left( {u{u_{2x}} + u_x^2} \right)d - k{u_x}{u_{2x}} - e\left( {u{u_{3x}} + {u_x}{u_{2x}}} \right) + {u_t}=0,
\end{equation}
where $g,n,k,d$ and $e$ are real parameters. Equation \eqref{Ks} models the pressure waves in a liquid and gas bubbles mixture when the viscosity of liquid and the heat transfer are both considered. In \cite{mirza}, authors  have found the exact traveling-wave solutions of Eq. \eqref{Ks} by using the first integral method. We have also employed the first integral method to find exact traveling-wave solutions of Eq. \eqref{Ks} with the help of GiNaCDE and compare our result with the result in \cite{mirza}.

Here we have used the following initializations:

\begin{center}
	\begin{tabular}{|c|c|c|}
		\hline
		\textbf{F-expansion:}&  \textbf{mF-expansion:}& \textbf{FIM:} \\
		\hline
		\begin{minipage}{2.3in}
			\small
			\begin{verbatim}
				twcPhase=lst{lst{k_0,k_1},lst{0,0}};
				degAcoeff=
				lst{3,A_0,A_1,A_2,A_3};
				ASolve=true;
				positivePart=true; 
				negativePart=true;
				paraInDiffSolve=lst{};
				filename="KS_Fexp.txt";
			\end{verbatim}
		\end{minipage}&  	
		\begin{minipage}{2.3in}
			\small
			\begin{verbatim}
				twcPhase=lst{lst{k_0,k_1},lst{0,0}};
				degAcoeff=lst{2,A_0,A_1,A_2};
				ASolve=true;
				positivePart=true; 
				negativePart=true;
				paraInDiffSolve=lst{};
				filename="KS_mF.txt";
			\end{verbatim}
		\end{minipage}&  
		\begin{minipage}{2.3in}
			\small
			\begin{verbatim}
				twcPhase=lst{lst{k_0,k_1},lst{0,0}};
				paraInDiffSolve=lst{};
				filename="KS_FIM.txt";
			\end{verbatim}
		\end{minipage}\\
		\hline
	\end{tabular}
\end{center}

In all methods, GiNaCDE makes the traveling-wave transformation \eqref{KdVB_twtrans} 
on the Eq. \eqref{Ks} and then it integrates the transformed NLPDE one time. In the case of F-expansion and mF-expansion methods we do not assign any value to integrating constant $ic_1$, and $ic_1$ behaves as unknown. But in FIM, we assign the integrating constant $ic_1$ to $0$. In the case of FIM, the software automatically detects that $h(X)$ is not polynomial in $X$. To avoid such non-polynomial form of $h(X)$, we have implemented the procedure explained in \cite{mirza} in step FIM4 of our algorithm. Following step FIM4, GiNaCDE performs a transformation to avoid singularity temporarily and the corresponding part of output where GiNaCDE makes the transformation is:
\begin{verbatim}
	We make the transformation, d xi = (-1+e*X_)*d eta to avoid singularity -1+e*X_ = 0 temporarily.
	Let U = X_, Diff(U,eta, 1) = Y_*(-1+e*X_), then we get
	Diff(X_,eta, 1) = Y_*(-1+e*X_),
	Diff(Y_,eta, 1) = -1/2*k*Y_^2-(d*k_1^(-1)*X_+n*k_1^(-1))*Y_+1/2*g*k_1^(-2)*X_^2+k_0*k_1^(-3)*X_,
\end{verbatim}
After that transformation, assuming $a_1=1$, the software evaluates $h=-\frac{k}{2}$. Then the degrees of $a_0,g$ are auto-evaluated following the strategy in step FIM5 and it finds two sets of balanced degrees which are $deg(a_0,g)=(2,0),(1,1)$. Second set of balanced degrees was also obtained in \cite{mirza}.  For first balanced degrees set $(2,0)$, the software finds the exact solutions of Eq. \eqref{Ks} only for $k_0=0,k_1=0$.
Interestingly all our results derived by GiNaCDE using FIM match with \cite{mirza} if we replace $k_0,k_1$ with $-c,1$ respectively.

Finally, we have also solved the above-listed five examples with the help of RAEEM Maple package \cite{raeem}. We have seen that RAEEM with Maple v. 8.0 is unable to solve the NLPDEs \eqref{eckhaus}, \eqref{gKdV} and \eqref{nlseWtKerr}. RAEEM can only solve the NLPDEs \eqref{sSK}, \eqref{Ks}, and gives 6, 6 solutions with CPU times 61 s, 250 s respectively. Beside the above examples, we have successfully solved more than 20 NLPDEs (provided in the result and test folders of GiNaCDE source code) using the GiNaCDE library and it gives results not more than 100 S.

\section{Conclusions}\label{sec:conc}
We have presented the algorithms for the high-performance automated F-expansion and First Integral Methods. We have also implemented these algorithms in a C++ library named GiNaCDE. This library is used to find the closed-form traveling-wave solution of the NLPDEs of the form \eqref{geneq}. The solution methods described in the Secs. \ref{sec:Fexpn}, \ref{sec:mF} and \ref{sec:fim}, are very tedious if we apply these three methods on an NLPDE by hand. The program library automates the methods and delivers possible solutions when they exist. However, in this context, in order to start the solution process, we have to provide some initial data (the options and parameters specified by the user) to the library. After running the solution process, there is also scope to assign the values to integrating constants generated from integrable NLPDEs. From these points of view, we can tell that our algorithms are not fully automated. But, these features make the library more efficient and powerful to tackle a large class of NLPDEs. Due to the implementations of three different methods in one software, one can easily check exact traveling-wave solutions of a huge variant of NLPDEs by applying those methods in one place with less labor.

We have introduced an exciting feature in our proposed algorithms, by which one can solve the complex NLPDEs. 
The program library can integrate the input-NLPDE or the transformed NLODE when possible. The generated integrating constants can be assigned with the values in our choice after running the program. Following \cite{mirza}, we have added another interesting feature in the algorithm for the first integral method. According to this new feature, our proposed algorithm is capable of tackling the non-polynomial form of $h (X)$ in Eq. \eqref{hform}.  
We have applied the package to a wide class of nonlinear evolution equations. It successfully recovered all previously known solutions that many powerful methods had found. More importantly, we have found new solutions and a more general form of solutions for some of the equations considered. 

\par Many computer packages such as RATH \cite{rath}, { \em PDESpecialSolutions.m} \cite{pdespclpkg, pdespclpkg1}, RAEEM \cite{raeem} are currently available which can solve the NLPDEs of the form \eqref{geneq} using some popular methods other than F-expansion, modified F-expansion, and first integral methods. These currently available packages cannot solve the NLPDEs containing complex conjugate functions, but GiNaCDE can solve such type of NLPDEs. We have compared the performances of GiNaCDE and RAEEM \cite{raeem} by solving some NLPDEs listed in Sec. \ref{sec:examp}. We have observed that RAEEM cannot solve $3$ NLPDEs among $5$ examples, but interestingly GiNaCDE can solve all the NLPDEs. We have also noted that GiNaCDE is much faster than RAEEM. We have tested the GiNaCDE library on more than 20 NLPDEs and it gives results not more than 100 S.   

\par The program library shows the results with calculating steps, and the results are saved in a text file. 
We have also provided a GUI version of our library that can be used without any programming knowledge. Thus, the GiNaCDE can be easily used to solve a broad class of NLPDEs for obtaining exact solutions of NLPDEs, and it is very efficient and fast to get the solutions.

Since our software performs the computations automatically from start to finish without human intervention except for few steps in the algorithms, it may not return the solutions in the simplest form. If required, the user can further manipulate and graph the solutions. The most important and vital step in the algorithms is analyzing and solving the nonlinear algebraic system. In the algorithms, most of the computation time is spent in solving the nonlinear algebraic system. So for a complicated algebraic system, the library takes more time to solve an NLPDE. Furthermore, sometimes the nonlinear algebraic system may be quintic or higher degree and may contain a large number of parameters, and then it may be unsolvable in analytic form. Therefore, although unlikely, due to the limitations of the algebraic solver, some exact solutions of NLPDEs may be missed.  

\par The GiNaCDE library for its algebraic manipulations depends only on the GiNaC library \cite{ginac}. GiNaC algebra system assigns a symbol id to each name of a symbol, and unlike other algebra systems, it uses the symbol id instead of its name for algebraic manipulations. Since the symbols id may change for each running session of the program, the same algebraic expressions may appear in different order of symbols. We have noted that the present version of GiNaCDE cannot handle a system of NLPDEs. However, it is expected that these limitations can be removed in the future version of GiNaCDE. For F-expansion and modified F-expansion methods, the solutions were considered in the form \eqref{soluseries}. Many researchers have considered the solutions in several other forms. Z. Sheng \cite{furimprfexpn}  considers solutions of the form
\begin{equation}\label{soluseries1}
	U = {a_0}(\xi ) + \sum\limits_{i = 1}^N {\left\{ {{a_{ - i}}(\xi ){F^{ - i}}(\xi ) + {a_i}(\xi ){F^i}(\xi ) + {b_i}(\xi ){F^{i - 1}}(\xi )\frac{{dF(\xi )}}{{d\xi }}} \right\}}.
\end{equation}
Here the parameters $a_0,a_i,a_{-i}$ and $b_i$ are not constant, and they depend on traveling-wave coordinate $\xi$. 
Y.M. Zhao \cite{extfexpn4} seeks the solutions of the form
\begin{equation}\label{soluseries2}
	U = {a_0} + \sum\limits_{i = 1}^N {\left( {{a_i}{F^i}(\xi ) + {b_i}{F^{i - 1}}(\xi )\frac{{dF(\xi )}}{{d\xi }}} \right)}.
\end{equation}
So there is a scope to extend the capabilities within GiNaCDE to find the solutions in the forms \eqref{soluseries1}, \eqref{soluseries2}. We will introduce such extensions in a future version of GiNaCDE.

\appendix

\renewcommand{\theequation}{A-\arabic{equation}}
\section{Solutions of Riccati equation}\label{sec:appenA}

	In case of Riccati equation, Eq. \eqref{1stnlode2} take the form
\begin{equation}\label{riccati}
	F'\left( \xi  \right) = {A_0} + {A_1}F + {A_2}{F^2}.
\end{equation} 

The solutions of the equation \eqref{riccati} are \cite{yang}
\begin{subequations}\label{riccati_solu}
	\begin{align}
		\label{riccati_solu1}F\left( \xi  \right) &= - \frac{{{A_1}}}{{2{A_2}}} - \frac{S}{{2{A_2}}}\tanh \left( {\frac{S}{2}\xi  + C} \right) \;\;(\text{If } A_2 \neq 0 \text{ and } A_1 \text{or} A_0 \neq 0),\\
		F\left( \xi  \right) &= - \frac{{{A_1}}}{{2{A_2}}} - \frac{S}{{2{A_2}}}\coth \left( {\frac{S}{2}\xi  + C} \right) \;\;(\text{If } A_2 \neq 0 \text{ and } A_1 \text{or} A_0 \neq 0),\\
		F\left( \xi  \right) &= {\left( { - \frac{{{A_1}}}{{2{A_0}}} + \frac{S}{{2{A_0}}}\tanh \left( {\frac{S}{2}\xi  + C} \right)} \right)^{ - 1}} \;\;(\text{If }A_0 \neq 0 \text{ and } A_1 \text{or} A_2 \neq 0),\\
		F\left( \xi  \right) &= {\left( { - \frac{{{A_1}}}{{2{A_0}}} + \frac{S}{{2{A_0}}}\coth \left( {\frac{S}{2}\xi  + C} \right)} \right)^{ - 1}} \;\;(\text{If }A_0 \neq 0 \text{ and } A_1 \text{or} A_2 \neq 0),\\
		\label{riccati_solu2e}F\left( \xi  \right) &=  - \frac{{{A_1}}}{{2{A_2}}} - \frac{S }{{2{A_2}}}\tanh \left( {\frac{{S \xi }}{2}+C} \right) + \frac{{\operatorname{sech}^2 \left( {\frac{{S \xi }}{2}+C} \right)}}{{{C} - \frac{{2{A_2}}}{S }\tanh \left( {\frac{{S \xi }}{2}+C} \right)}}\\
		& \,\,\,\,\,\,\,\,\,\,\,\,\,\,\,\,\,\,\,\,\,\,\,\,\,\,\,\,\,\,(\text{If } A_2 \neq 0 \text{ and } A_1 \text{or} A_0 \neq 0),\nonumber\\
		\label{riccati_soluA2f}F\left( \xi  \right) &= {\text{ - }}\frac{{{A_0}}}{{{A_1}}} + C{e^{ {{A_1}\xi }}},\;\; (\text{If } A_2=0 \text{ and } A_1\neq 0),\\
		F\left( \xi  \right) &= A_0\xi +C\;\;(\text{If }  A_2=A_1=0 \text{ and } A_0\neq 0 ),	
	\end{align}
\end{subequations}

where $S = \sqrt{A_1^2-4A_0A_2}$ and $C$ is auxiliary constant. In the above solutions, from Eq. \eqref{riccati_solu1} to \eqref{riccati_solu2e}, the condition $A_1^2-4A_0A_2>0$ must be satisfied. 

Please note that in the following solutions if not mentioned $C$ has to be assumed as an auxiliary constant.

\renewcommand{\theequation}{B-\arabic{equation}}
\section{Solutions of Bernoulli equation}\label{sec:appenB}

In case of Bernoulli equation, Eq. \eqref{1stnlode2} is reduced to
\begin{equation}\label{bernouli}
F'\left( \xi  \right) = {A_1}F + {A_\delta}F^{\delta}.
\end{equation} 

The solutions of the equation \eqref{bernouli} are
\begin{subequations}\label{bernouli_solu}
\begin{align}
F\left( \xi  \right) &= {\left( {\frac{{{A_1} \left( {\cosh \left( {{A_1}\left( {\delta - 1} \right) \xi + {C} {A_1}} \right) + \sinh \left( {{A_1} \left( {\delta - 1} \right)\xi + {C} {A_1}} \right)} \right)}}{{1 - {A_\delta} \cosh \left( {{A_1}\left( {\delta - 1} \right) \xi + {C} {A_1}} \right) - {A_\delta} \sinh \left( {{A_1}\left( {\delta - 1} \right) \xi + {C} {A_1}} \right)}}} \right)^{{{\left( {\delta - 1} \right)}^{ - 1}}}} \;\;(\text{If } A_1\neq 0\text{ and }\delta \neq 1),\\
F\left( \xi  \right) &= {\left( { - \frac{{{A_\delta }}}{{{A_1}}} + C{e^{{A_1}(1 - \delta )\xi }}} \right)^{ \frac{1}{{1 - \delta }}}}\;\;(\text{If } A_1\neq 0\text{ and }\delta \neq 1), \\
F\left( \xi  \right) &= {\left( { - \frac{{{A_1}}}{{2{A_\delta }}} - \frac{{{A_1}}}{{2{A_\delta }}}\tanh \left( {\frac{{(\delta  - 1){A_1}}}{2}\xi  + C} \right)} \right)^{\frac{1}{{\delta  - 1}}}}\;\;(\text{If } A_1\neq 0\text{ and }\delta \neq 1),\\
F\left( \xi  \right) &= {\left( { - \frac{{{A_1}}}{{2{A_\delta }}} - \frac{{{A_1}}}{{2{A_\delta }}}\coth \left( {\frac{{(\delta  - 1){A_1}}}{2}\xi  + C} \right)} \right)^{\frac{1}{{\delta  - 1}}}}\;\;(\text{If } A_1\neq 0\text{ and }\delta \neq 1), \\ 
F\left( \xi  \right) &= {\left( {{{\left( {{A_\delta} \xi(1 -  \delta) + {C}} \right)}^{{{\left( {\delta - 1} \right)}^{ - 1}}}}} \right)^{ - 1}} \;\;(\text{If } A_1=0 \text{ and } \delta \neq 1)),\\
F\left( \xi  \right) &= C e^{({A_1} + {A_\delta })\xi }\;\; (\text{If }\delta = 1).
\end{align}
\end{subequations}

\renewcommand{\theequation}{C-\arabic{equation}}
\section{first-order NLODEs related to Jacobi Elliptic Functions}\label{sec:appenC}
The incomplete elliptic integral of the first kind, is defined by
\begin{equation}\label{ellipInt}
u(\phi,m) = \int\limits_0^\phi  {\frac{{d\theta }}{{\sqrt {1 - {m^2}{{\sin }^2}\theta } }}},
\end{equation}
where $m$ is the elliptic modulus, and $\phi=\text{JacobiAM}(u,m)$ is the Jacobi amplitude which is the inverse of  elliptic integral \eqref{ellipInt}. The three principal elliptic functions are denoted $\text{JacobiSN}(u,m),\text{JacobiCN}(u,m),\text{JacobiDN}(u,m),$ which are in turn defined in terms of the amplitude function JacobiAM satisfying
\begin{align}
&\text{JacobiSN}(u,m)=\sin\left(\text{JacobiAM}(u,m)\right),\\
&\text{JacobiCN}(u,m)=\cos\left(\text{JacobiAM}(u,m)\right),\\
&\text{JacobiDN}(u,m)=\frac{\partial}{\partial u}\text{JacobiAM}(u,m)=\sqrt{1-m^2\text{JacobiSN}(u,m)^2}.
\end{align}
There are total twelve Jacobian functions that can be expressed in general by a name $\text{JacobiXY}$ which follows the identities $\text{JacobiXY}=\frac{1}{\text{JacobiYX}}=\frac{\text{JacobiPR}}{\text{JacobiQR}}$. Here $X,Y,R$ are any three of $S,C,N,D$. Following these rules of notations other nine subsidiary Jacobian elliptic functions can be defined in terms of the three JacobiSN, JacobiCN, JacobiDN by the following identities:
\begin{align}
&  {\text{JacobiNS}}(u,m) = {({\text{JacobiSN}}(u,m))^{ - 1}},{\text{JacobiND}}(u,m) = {({\text{JacobiDN}}(u,m))^{ - 1}},{\text{JacobiSC}}(u,m) = {({\text{JacobiCS}}(u,m))^{ - 1}}, \hfill\nonumber \\
& {\text{JacobiSD}}(u,m) = {({\text{JacobiDS}}(u,m))^{ - 1}},{\text{JacobiDC}}(u,m) = {({\text{JacobiCD}}(u,m))^{ - 1}}, {\text{JacobiNC}}(u,m) = {({\text{JacobiCN}}(u,m))^{ - 1}},\nonumber\\
&{\text{JacobiCS}}(u,m) = \frac{{{\text{JacobiCN}}(u,m)}}{{{\text{JacobiSN}}(u,m)}},{\text{JacobiDS}}(u,m) = \frac{{{\text{JacobiDN}}(u,m)}}{{{\text{JacobiSN}}(u,m)}},{\text{JacobiCD}}(u,m) = \frac{{{\text{JacobiCN}}(u,m)}}{{{\text{JacobiDN}}(u,m)}}.
\end{align}
Jacobian elliptic functions can also be defined as solutions to the differential equations
\begin{equation}\label{024}
F'(\xi)=\sqrt{A_0+A_2F^2+A_4F^4}.
\end{equation}
 Solutions of Eq. \eqref{024} in terms of  Jacobi elliptic functions are listed in below (with the conditions that all the algebraic expressions within square-root must be greater than $0$):
 \begin{subequations}\label{024_solu}
\begin{align}
&F(\xi)=   \frac{{\sqrt {2{A_2} {A_4} \left( { - {A_2}^2 + \sqrt {{A_2}^2S} } \right)} }}{{2{A_2} {A_4}}}{\text{JacobiSN}}\left( {\frac{{\sqrt {2{A_2} \left( { - {A_2}^2 - \sqrt {{A_2}^2S} } \right)} \xi }}{{2{A_2}}},\frac{{\sqrt {2{A_0} {A_4} \left( { - 2 {A_0} {A_4} + {A_2}^2 - \sqrt {{A_2}^2S} } \right)} }}{{2{A_0} {A_4}}}} \right), \hfill \\
&F(\xi)=    \frac{{\sqrt {2{A_2} {A_4} \left( { - {A_2}^2 + \sqrt {{A_2}^2S} } \right)} }}{{2{A_2} {A_4}}}{\text{JacobiCN}}\left( {\frac{{\sqrt { - {A_2} \sqrt {{A_2}^2S} } \xi }}{{{A_2}}},\frac{{\sqrt { - 2S\left( {\sqrt {{A_2}^2S}  - S} \right)} }}{{2S}}} \right), \hfill \\
&F(\xi)=    \frac{{\sqrt {2{A_2} {A_4} \left( { - {A_2}^2 - \sqrt {{A_2}^2S} } \right)} }}{{2{A_2} {A_4}}}{\text{JacobiDN}}\left( { \frac{{\sqrt {2{A_2} \left( {{A_2}^2 + \sqrt {{A_2}^2S} } \right)} \xi }}{{2{A_2}}}, \frac{{\sqrt {2{A_0} {A_4} \left( {\sqrt {{A_2}^2S}  - S} \right)} }}{{2{A_0} {A_4}}}} \right), \hfill \\
&F(\xi)=    \frac{{\sqrt {2{A_4} \left( { - {A_2} + \sqrt S } \right)} }}{{2{A_4}}}{\text{JacobiNS}}\left( {1/2 \sqrt { - 2 {A_2} + 2 \sqrt S } \xi , \frac{{\sqrt {2{A_0} {A_4} \left( {{A_2} \sqrt S  - 2 {A_0} {A_4} + {A_2}^2} \right)} }}{{2{A_0} {A_4}}}} \right), \hfill \\
&F(\xi)=   \frac{{\sqrt { - 2{A_2} - 2\sqrt S } }}{{2\sqrt {{A_4}} }}{\text{JacobiNC}}\left( {\sqrt { - \sqrt S } \xi ,\frac{{\sqrt { - 2 {A_0} {A_4}} }}{{\sqrt {{A_2} \sqrt S  + S} }}} \right),\\
&F(\xi)=   \frac{{\sqrt { - 2{A_2} + 2\sqrt S } }}{{2\sqrt {{A_4}} }}{\text{JacobiND}}\left( {\frac{{\sqrt { - 2 {A_0} {A_4}} \xi }}{{\sqrt { - {A_2} + \sqrt S } }},\frac{{\sqrt {2{A_2} \sqrt S  - 2S} }}{{2\sqrt {{A_0} {A_4}} }}} \right), \hfill \\
&F(\xi)=   - \frac{{\sqrt 2 \sqrt {{A_0}} }}{{\sqrt {{A_2} + \sqrt S } }}{\text{JacobiSC}}\left( {\frac{{\sqrt {{A_2} \sqrt S  - 2 {A_0} {A_4} + {A_2}^2} \xi }}{{\sqrt {{A_2} + \sqrt S } }},\frac{{\sqrt {{A_2} \sqrt S  + S} }}{{\sqrt {{A_2} \sqrt S  - 2 {A_0} {A_4} + {A_2}^2} }}} \right), \hfill \\
&F(\xi)=  \frac{{\sqrt 2 \sqrt {{A_0}} \sqrt { - {A_2} \sqrt S  + 2 {A_0} {A_4} - {A_2}^2} }}{{\sqrt {{A_2} + \sqrt S } \sqrt { - {A_2} \sqrt S  - S} }}{\text{JacobiSD}}\left( {\frac{{\sqrt {{A_2} \sqrt S  + S} \xi }}{{\sqrt {{A_2} + \sqrt S } }},\frac{{\sqrt { - {A_2} \sqrt S  + 2 {A_0} {A_4} - {A_2}^2} }}{{\sqrt { - {A_2} \sqrt S  - S} }}} \right), \hfill \\
&F(\xi)=  \frac{{\sqrt 2 \sqrt {{A_0}} }}{{\sqrt {{A_2} + \sqrt S } }}{\text{JacobiCS}}\left( {\frac{{ \sqrt {2{A_0} {A_4}} \xi }}{{\sqrt {{A_2} + \sqrt S } }},\frac{{\sqrt { - {A_2} \sqrt S  - S} }}{{\sqrt {2{A_0} {A_4}} }}} \right), \hfill \\
&F(\xi)=  \frac{{\sqrt {{A_2} \sqrt S  - S} }}{{\sqrt {{A_2} {A_4} - {A_4} \sqrt S } }}{\text{JacobiDS}}\left( {\frac{{\sqrt {{A_2} \sqrt S  - S} \xi }}{{\sqrt {{A_2} - \sqrt S } }},\frac{{\sqrt {2{A_0} {A_4}} }}{{\sqrt {{A_2} \sqrt S  - S} }}} \right),
\end{align}
\end{subequations}
where $S = \left( { - 4 {A_0} {A_4} + {A_2}^2} \right)$. We note that Jacobian functions in all the above solutions are in the form $C_1\text{JacobiXY}(C_2\xi,C_3)$, where $C_1,C_2,C_3$ are constants which contains only the constant parameters of differential equation \eqref{024}. It will be better to express them in the convenient form $\text{JacobiXY}(\xi,m)$ by putting $C_1=1,C_2=1,C_3=m$. We derive the values of the constant parameters $A_0,A_2,A_4$ in terms of modulus $m$ by solving the system of equations $\{C_1=1,C_2=1,C_3=m\}$ simultaneously. There have some advantages to have solutions in the form  $\text{JacobiXY}(\xi,m)$. We can transform $\text{JacobiSN}(\xi,m),\text{JacobiCN}(\xi,m), \text{JacobiDN}(\xi,m)$ into hyperbolic functions $\tanh(\xi),\operatorname{sech}(\xi),\operatorname{sech}(\xi)$ respectively simply making the approximation $m\rightarrow 1$. We can also simply get the trigonometric transformations $\text{JacobiSN}(\xi,m)\rightarrow\sin(\xi), \text{JacobiCN}(\xi,m)\rightarrow\cos (\xi)$ and also get the transformation $\text{JacobiDN}(\xi,m)\rightarrow1$ in the limit case $m\rightarrow 0$. So by choosing a single A.E. \eqref{024} one can obtain a variety of exact solutions of NLPDEs in terms of Jacobian elliptic functions, hyperbolic functions and trigonometric functions. Table \ref{jefSpsolu} shows all the solutions of the parameters $A_0,A_2,A_4$ solving the system of equations $\{C_1=1,C_2=1,C_3=m\}$ and the corresponding solutions of Eq. \eqref{024} are given in last column. Interestingly the values of $A_0,A_2,A_4$ in Table \ref{jefSpsolu} coincide with the ones given in \cite{fexpn}. 
\begin{table}[H]
\centering
\caption{ Solutions $F(\xi)$ of Eq. \eqref{024} in terms of Jacobi elliptic functions for some values of $A_0$,$A_2$ and $A_4$ \cite{fexpn}. Here the values of $A_0$,$A_2$ and $A_4$ are taken through modulus ($m$) in most cases.}
\begin{tabular}{c@{\hskip .0in}c@{\hskip .3in}c@{\hskip .3in}c@{\hskip .3in}c}
 & $A_0$ & $A_2$ & $A_4$ & $F(\xi)$ \\ [0.5ex]
\hline \hline\\
& $1$ & $-(1+m^2)$ & $m^2$ & $\text{JacobiSN}(\xi,m)$\\[0.5ex]
&  $1-m^2$ & $2m^2-1$ & $-m^2 $ & $\text{JacobiCN}(\xi,m) $ \\[0.5ex]
&  $m^2-1$ & $2-m^2$ & $-1$ & $\text{JacobiDN}(\xi,m)$ \\[0.5ex]
&  $m^2$ & $-(1+m^2)$ & $1$ & $\text{JacobiNS}(\xi,m)=(\text{JacobiSN}(\xi,m))^{-1}$ \\[0.5ex]
&  $-m^2$ & $2m^2-1$ & $1-m^2$ & $\text{JacobiNC}(\xi,m)=(\text{JacobiCN}(\xi,m))^{-1}$ \\[0.5ex]
&  $-1$ & $2-m^2$ & $m^2-1$ & $\text{JacobiND}(\xi,m)=(\text{JacobiDN}(\xi,m))^{-1}$ \\[0.5ex]
&  $1$ & $2-m^2$ & $1-m^2$ & $\text{JacobiSC}(\xi,m)=\frac{\text{JacobiSN}(\xi,m)}{\text{JacobiCN}(\xi,m)}$ \\[0.5ex]
&  $1$ & $2m^2-1$ & $-m^2(1-m^2)$ & $\text{JacobiSD}(\xi,m)=\frac{\text{JacobiSN}(\xi,m)}{\text{JacobiDN}(\xi,m)}$ \\[0.5ex]
&  $1-m^2$& $2-m^2$ & $1$  & $\text{JacobiCS}(\xi,m)=\frac{\text{JacobiCN}(\xi,m)}{\text{JacobiSN}(\xi,m)}$ \\[0.5ex]
&  $-m^2(1-m^2)$ & $2m^2-1$ & $1$ & $\text{JacobiDS}(\xi,m)=\frac{\text{JacobiDN}(\xi,m)}{\text{JacobiSN}(\xi,m)}$ \\ 
 [1ex] 
\hline 
\end{tabular}
\label{jefSpsolu}
\end{table}

 The solutions of an another first-order NLODE
\begin{equation}\label{123}
F'(\xi)=\sqrt{A_1F+A_2F^2+A_3F^3}, 
\end{equation}
can also be expressed in terms of Jacobi elliptic functions. The solutions of Eq. \eqref{123} are (with the conditions that all the algebraic expressions within square-root must be greater than $0$)
\begin{subequations}\label{123_solu}
\begin{align}
&F(\xi)=    \frac{{ - {A_2} + \sqrt S }}{{2{A_3}}}{\text{JacobiS}}{{\text{N}}^2}\left( { \frac{{\sqrt {2{A_1} {A_3}} \xi }}{{2\sqrt { - {A_2} + \sqrt S } }}, \frac{{\sqrt {{A_2}^2 - {A_2} \sqrt S  - 2 {A_1} {A_3}} }}{{\sqrt {2{A_1} {A_3}} }}} \right), \hfill \\
&F(\xi)=    -  \frac{{2{A_1}}}{{{A_2} - \sqrt S }}{\text{JacobiC}}{{\text{N}}^2}\left( { \frac{{\sqrt[4]{S}\xi }}{2}, \frac{{\sqrt {{A_2} + \sqrt S } }}{{\sqrt 2 \sqrt[4]{S}}}} \right), \hfill \\
&F(\xi)=    - \frac{{2{A_1}}}{{{A_2} - \sqrt S }}{\text{JacobiD}}{{\text{N}}^2}\left( {\frac{{\sqrt {2{A_1} {A_3}} \xi }}{{2\sqrt {{A_2} - \sqrt S } }}, \frac{{\sqrt {4 {A_1} {A_3} + \sqrt S {A_2} - {A_2}^2} }}{{\sqrt {2{A_1} {A_3}} }}} \right), \hfill \\
&F(\xi)=    - \frac{{2{A_1}}}{{{A_2} + \sqrt S }}{\text{JacobiN}}{{\text{S}}^2}\left( { \frac{{\sqrt 2 }}{4}\sqrt { - {A_2} + \sqrt S } \xi ,\frac{{\sqrt {{A_2} + \sqrt S } }}{{\sqrt {{A_2} - \sqrt S } }}} \right), \hfill \\
&F(\xi)=    - \frac{{2{A_1}}}{{{A_2} + \sqrt S }}{\text{JacobiN}}{{\text{C}}^2}\left( { \frac{{\sqrt[4]{S}\xi }}{2},\frac{{\sqrt {{A_2} + \sqrt S } }}{{\sqrt 2 \sqrt[4]{S}}}} \right), \hfill \\
&F(\xi)=    - \frac{{2{A_1}}}{{{A_2} + \sqrt S }}{\text{JacobiN}}{{\text{D}}^2}\left( { \frac{{\sqrt 2 \sqrt {{A_2} + \sqrt S } \xi }}{4},\frac{{\sqrt 2 \sqrt[4]{S}}}{{\sqrt {{h_2} + \sqrt S } }}} \right), \hfill \\
&F(\xi)=   \frac{{2{A_1}}}{{{A_2} + \sqrt S }}{\text{JacobiS}}{{\text{C}}^2}\left( {\frac{{\sqrt {\sqrt S {A_2} - 2 {A_1} {A_3} + {A_2}^2} \xi }}{{2\sqrt {{A_2} + \sqrt S } }},\frac{{\sqrt {\sqrt S {A_2} + S} }}{{\sqrt {\sqrt S {A_2} - 2 {A_1} {A_3} + {A_2}^2} }}} \right), \hfill \\
&F(\xi)=   \frac{{2{A_1} \left( { - 2 {A_1} {A_3} + \sqrt S {A_2} + {A_2}^2} \right)}}{{\left( {{A_2} + \sqrt S } \right)\left( {\sqrt S {A_2} + S} \right)}}{\text{JacobiS}}{{\text{D}}^2}\left( {\frac{{\sqrt {\sqrt S {A_2} + S} \xi }}{{2\sqrt {{A_2} + \sqrt S } }},\frac{{\sqrt {2 {A_1} {A_3} - \sqrt S {A_2} - {A_2}^2} }}{{\sqrt { - \sqrt S {A_2} - S} }}} \right), \hfill \\
&F(\xi)=   \frac{{2{A_1}}}{{{A_2} - \sqrt S }}{\text{JacobiC}}{{\text{S}}^2}\left( {\frac{{\sqrt {{A_1} {A_3}} \xi }}{{\sqrt 2 \sqrt {{A_2} - \sqrt S } }}, \frac{{\sqrt {\sqrt S {A_2} - S}  }}{{\sqrt {2{A_1} {A_3}} }}} \right), \hfill \\
&F(\xi)=   \frac{{ - \sqrt S {A_2} - S}}{{{A_3}\left( {{A_2} + \sqrt S } \right)}}{\text{JacobiD}}{{\text{S}}^2}\left( {\frac{{\sqrt { - \sqrt S {A_2} - S} \xi }}{{2\sqrt {{A_2} + \sqrt S } }},\frac{{ \sqrt {2{A_1} {A_3}} }}{{\sqrt { - \sqrt S {A_2} - S} }}} \right),
\end{align}
\end{subequations}
where
\begin{align}  
&S =  - 4 {A_1} {A_3} + {A_2}^2.
\end{align}

Using the values of $A_1,A_2,A_3$ listed in Table \ref{jefSpsolu2}, one can transforms Jacobi elliptic functions which are present in the solutions into the simplified form $\text{JacobiXY}(\xi,m)$. 

\begin{table}[H]
\centering
\caption{ Solutions $F(\xi)$ of Eq. \eqref{123} in terms of Jacobi elliptic functions for some values of $A_1$,$A_2$ and $A_3$. Here the values of $A_1$,$A_2$ and $A_3$ are taken through modulus ($m$) in most cases.}
\begin{tabular}{c@{\hskip .0in}c@{\hskip .3in}c@{\hskip .3in}c@{\hskip .3in}c}
 & $A_1$ & $A_2$ & $A_3$ & $F(\xi)$ \\ [0.5ex]
\hline \hline\\
& $4$ & $-4(1+m^2)$ & $4m^2$ & $\text{JacobiSN}^2(\xi,m)$\\[0.5ex]
&  $-4(m^2-1) $ & $4(2m^2-1)$ & $-4m^2$ & $\text{JacobiCN}^2(\xi,m) $ \\[0.5ex] 
&  $4(m^2-1)$ & $4(2-m^2)$ & $-4$ & $\text{JacobiDN}^2(\xi,m)$ \\
&  $4m^2$ & $-4(1+m^2)$ & $4$ & $\text{JacobiNS}^2(\xi,m)$ \\[0.5ex] 
&  $-4m^2$ & $4(2m^2-1)$ & $-4(m^2-1)$ & $\text{JacobiNC}^2(\xi,m)$ \\[0.5ex] 
&  $-4$ & $4(2-m^2)$ & $4(m^2-1)$ & $\text{JacobiND}^2(\xi,m)$ \\[0.5ex] 
&  $4$ & $4(2-m^2)$ & $4(1-m^2)$ & $\text{JacobiSC}^2(\xi,m)$ \\[0.5ex] 
&  $4$ & $4(2m^2-1)$ & $-4m^2(1-m^2)$ & $\text{JacobiSD}^2(\xi,m)$ \\[0.5ex] 
&  $4(1-m^2)$ & $4(2-m^2)$ & $4$ & $\text{JacobiCS}^2(\xi,m)$ \\[0.5ex] 
&  $-4m^2(1-m^2)$ & $4(2m^2-1)$ & $4$ & $\text{JacobiDS}^2(\xi,m)$ \\ 
 [1ex] 
\hline 
\end{tabular}
\label{jefSpsolu2}
\end{table}
\renewcommand{\theequation}{D-\arabic{equation}}
\section{Solutions of some more different types of first-order NLODEs}\label{sec:appenD}
In this section, we discuss the solutions of more different types of the first-order NLODE. Among them, we can choose an auxiliary equation suitable for input-NLPDE.\\ 
\textbf{Type-1 first-order NLODE:}
\par Let us first consider the first-order NLODE as follows 
\begin{equation}\label{0246}
	F'(\xi ) = \sqrt{{A_0} + {A_2}{F^2} + {A_4}{F^4} + {A_6}{F^6}}.
\end{equation}
The above equation admits following special hyperbolic solutions \cite{0246}:

If ${A_0} = \frac{{8A_2^2}}{{27{A_4}}}\,{\text{and}}\,{A_6} = \frac{{A_4^2}}{{4{A_2}}}$ then it has a bell profile solution
\begin{subequations}\label{0246_solu}
	\begin{align}
		F(\xi) &= {\left( { -  \frac{{8{A_2} {{\left( {\tanh \left( { \sqrt { -  \frac{{{A_2}}}{3}} \xi  + C} \right)} \right)}^2}}}{{3{A_4} \left( {3 + {{\left( {\tanh \left( {\sqrt { -  \frac{{{A_2}}}{3}} \xi  + C} \right)} \right)}^2}} \right)}}} \right)^{\frac{1}{2}}}, \hfill \\
		\intertext{and a singular solution}\;\;\;\;\;\;\;\;\;\;\;\;\;\;\;\;\;\;\;\;\;\;\;\;\;\;\;\;\;\;\;\;\;\;\;\;\;\;\;\;\;\;\;\;\nonumber\\
		F(\xi) &=  {\left( { -  \frac{{8{A_2} {{\left( {\coth \left( { \sqrt { -  \frac{{{A_2}}}{3}} \xi  + C} \right)} \right)}^2}}}{{3{A_4} \left( {3 + {{\left( {\coth \left( {\sqrt { -  \frac{{{A_2}}}{3}} \xi  + C} \right)} \right)}^2}} \right)}}} \right)^{\frac{1}{2}}}\,.
	\end{align}
\end{subequations}
\\ 
\textbf{Type-2 first-order NLODE:}
\par For $A_0=0$ Eq. \eqref{0246} is reduced to
\begin{equation}\label{246}
	F'(\xi ) = \sqrt{{A_2}{F^2} + {A_4}{F^4} + {A_6}{F^6}}.
\end{equation}
Equation \eqref{246} has a triangular periodic solution \cite{0246}
\begin{subequations}\label{246_solu1}
	\begin{equation}\label{246_1}
		F(\xi ) = {\left( {\frac{{2{A_2}{{\operatorname{sech} }^2}(\sqrt {{A_2}} \xi  + C)}}{{2\sqrt {A_4^2 - 4{A_2}{A_6}}  - \left( {\sqrt {A_4^2 - 4{A_2}{A_6}}  + {A_4}} \right){{\operatorname{sech} }^2}(\sqrt {{A_2}} \xi  + C)}}} \right)^{\frac{1}{2}}},
	\end{equation}
	and a singular triangular periodic solution
	\begin{equation}\label{246_2}
		F(\xi ) = {\left( {\frac{{2{A_2}{{\operatorname{csch} }^2}(\sqrt {{A_2}} \xi  + C)}}{{2\sqrt {A_4^2 - 4{A_2}{A_6}}  + \left( {\sqrt {A_4^2 - 4{A_2}{A_6}}  - {A_4}} \right){{\operatorname{csch} }^2}(\sqrt {{A_2}} \xi  + C)}}} \right)^{\frac{1}{2}}}.
	\end{equation}
\end{subequations}
If ${A_6} = \frac{{A_4^2}}{{4{A_2}}}$, Eq. \eqref{246} also admits a kink profile solution
\begin{subequations}\label{246_solu2}
	\begin{equation}\label{246_3}
		F(\xi)={\left( { - \frac{{{A_2}}}{{{A_4}}} \left( {1 + \tanh \left( {\sqrt {{A_2}} \xi  + C} \right)} \right)} \right)^{\frac{1}{2}}},
	\end{equation}
	and a singular solution
	\begin{equation}\label{246_4}
		F(\xi)={\left( { - \frac{{{A_2}}}{{{A_4}}} \left( {1 + \coth \left( {\sqrt {{A_2}} \xi  + C} \right)} \right)} \right)^{\frac{1}{2}}}.
	\end{equation}
\end{subequations}
When $A_4=0$ Eq. \eqref{246} is reduced to
\begin{equation}\label{26}
	F'(\xi ) = \sqrt{{A_2}{F^2} + {A_6}{F^6}}.
\end{equation}
It is clear that the Eqs. \eqref{246_1}, \eqref{246_2} with $A_4=0$ are also solutions of Eq. \eqref{26}, but Eqs. \eqref{246_3}, \eqref{246_4} are not solutions of Eq. \eqref{26} as they are undefined at $A_4=0$.\\
\textbf{Type-3 first-order NLODE:}
\par An another type of first-order NLODE is
\begin{equation}\label{234}
	F'(\xi ) = \sqrt{{A_2}{F^2} + {A_3}{F^3} + {A_4}{F^4}},
\end{equation}  
which has different type solitary wave solutions \cite{234}
\begin{subequations}\label{234_solu}
	\begin{align}
		\label{234_1} F(\xi) &=   - \frac{{{A_2} {A_3} {{\left( {{\text{sech}}\left( { \frac{{\sqrt {{A_2}} }}{2}\xi } \right)} \right)}^2}}}{{{A_3}^2 - {A_2} {A_4} {{\left( {1 - \tanh \left( {\frac{{\sqrt {{A_2}} }}{2}\xi } \right)} \right)}^2}}}, \; F(\xi) =   \frac{{2{A_2} {\text{sech}}\left( {\sqrt {{A_2}} \xi } \right)}}{{\sqrt { - 4 {A_2} {A_4} + {A_3}^2}  - {A_3} {\text{sech}}\left( {\sqrt {{A_2}} \xi } \right)}}. 
	\end{align}
\end{subequations}
For $A_4=0$ Eq. \eqref{234} is reduced to
\begin{equation}\label{23}
	F'(\xi ) = \sqrt{{A_2}{F^2} + {A_3}{F^3}}.
\end{equation}
One can get the solutions of above equations substituting $A_4=0$ in solutions \eqref{234_1}, \eqref{234_2}. If $A_3=0$ Eq. \eqref{234} is simplified to 
\begin{equation}\label{24}
	F'(\xi ) = \sqrt{{A_2}{F^2} + {A_4}{F^4}}.
\end{equation}
Solutions \eqref{234_1}, \eqref{234_2} with $A_3=0$ also exist for \eqref{24}. Beside these solutions Eq. \eqref{24} has following two extra solutions:
\begin{subequations}\label{24_solu}
	\begin{align}
		F(\xi) &=   \frac{{4{A_2} {{\text{e}}^{\left( {\xi  + C} \right)\sqrt {{A_2}} }}}}{{ - 4 {A_2} {A_4} {{\text{e}}^{2 \sqrt {{A_2}} \xi }} + {{\text{e}}^{2 C \sqrt {{A_2}} }}}}, \;
		F(\xi) =   \frac{{4{A_2} {{\text{e}}^{\left( {\xi  + C} \right)\sqrt {{A_2}} }}}}{{ - 4 {A_2} {A_4} {{\text{e}}^{2 C \sqrt {{A_2}} }}{\text{ + }}{{\text{e}}^{2 \sqrt {{A_2}} \xi }}}}.
	\end{align}
\end{subequations}
\textbf{Type-4 first-order NLODE:}
\par We will now consider one more simplified first-order NLODE
\begin{equation}\label{02}
	F'(\xi ) = \sqrt{{A_0} + {A_2}{F^2}}.
\end{equation}
We find some exact solutions of Eq. \eqref{02} containing exponential, hyperbolic functions, which are listed below
\begin{subequations}\label{02_solu}
	\begin{align}
		F(\xi) &=  \frac{{\left( { - {A_0} {{\text{e}}^{2 \xi \sqrt {{A_2}} }} + {{\text{e}}^{2 C\sqrt {{A_2}} }}} \right){{\text{e}}^{ - \left( {C + \xi } \right)\sqrt {{A_2}} }}}}{{2\sqrt {{A_2}} }}, \;
		F(\xi) =  \frac{{\left( { -  {{\text{e}}^{2 \xi \sqrt {{A_2}} }} + {A_0}{{\text{e}}^{2 C\sqrt {{A_2}} }}} \right){{\text{e}}^{ - \left( {C + \xi } \right)\sqrt {{A_2}} }}}}{{2\sqrt {{A_2}} }}, \hfill \\
		F(\xi) &=   \pm \sqrt {\frac{{ - {A_0}}}{{{A_2}}}} \cosh \left( {\xi \sqrt {{A_2}}  + C} \right), \;
		F(\xi) =   \pm \sqrt {\frac{{{A_0}}}{{{A_2}}}} \sinh \left( {\xi \sqrt {{A_2}}  + C} \right). 
	\end{align}
\end{subequations}

\

\end{document}